\documentclass[
aps,groupedaddress,
amsfonts,amsmath,floatfix,eqsecnum,nofootinbib
,twocolumn
]{revtex4}

\usepackage[utf8]{inputenc}
\usepackage{epsfig,graphicx}
\usepackage{bm}
\usepackage{amsmath,amsfonts}
\usepackage{enumerate}
\usepackage{ytableau}
\usepackage{cancel}
\usepackage{multirow}
\usepackage{mathptmx}      
\usepackage{latexsym}
\usepackage{amstext}
\usepackage{array}
\usepackage[T1]{fontenc} 

\newcommand{\ignore}[1]{}
\newcommand{\nec}[1]{\eqref{eq:#1}}

\newcommand{\Eq}[1]{Eq.~(\ref{eq:#1})}
\newcommand{\Eqs}[1]{Eqs.~(\ref{eq:#1})}
\newcommand{\be}{\begin{equation}}
\newcommand{\ee}{\end{equation}}
\newcommand{\C}{\mathbb{C}}
\newcommand{\N}{\mathbb{N}}
\newcommand{\R}{\mathbb{R}}
\newcommand{\Z}{\mathbb{Z}}

\newcommand{\GL}{\mathrm{GL}}

\renewcommand{\S}{\mathrm{S}}
\newcommand{\SL}{\mathrm{SL}}

\newcommand{\SU}{\mathrm{SU}}
\newcommand{\U}{\mathrm{U}}

\newcommand{\su}{\mathrm{su}}

\newcommand{\mfootnote}[1]{\footnote{\sf #1}}

\renewcommand{\Re}[1]{{\,{\mathrm{Re}}\, #1}}
\renewcommand{\Im}[1]{{\,{\mathrm{Im}}\, #1}}

\def\slashchar#1{\setbox0=\hbox{$#1$}
   \dimen0=\wd0 \setbox1=\hbox{/} \dimen1=\wd1
   \ifdim\dimen0>\dimen1 \rlap{\hbox to \dimen0{\hfil/\hfil}} #1
   \else  \rlap{\hbox to \dimen1{\hfil$#1$\hfil}} / \fi}

\newcommand{\h}{{\textstyle{\frac{1}{2}}}}

\newcommand{\ben}{\begin{enumerate}}
\newcommand{\een}{\end{enumerate}}

\newcommand{\ds}{\displaystyle}

\newcommand{\tr}{{\mathrm{tr}}}

\newcommand{\sign}{\,{\mathrm{sign}}}
\def\diag{{\rm diag}}

\newcommand{\cA}{{\mathcal A }}

\newcommand{\cD}{{\mathcal D }}

\newcommand{\cM}{{\mathcal M }}
\newcommand{\cN}{{\mathcal N }}
\newcommand{\cO}{{\mathcal O }}

\newcommand{\cR}{{\mathcal R }}

\newcommand{\va}{{\bm a}}

\newcommand{\vh}{{\bm h}}
\newcommand{\vk}{{\bm k}}
\newcommand{\vn}{{\bm n}}
\newcommand{\vp}{{\bm p}}

\newcommand{\vx}{{\bm x}}
\newcommand{\vy}{{\bm y}}
\newcommand{\vz}{{\bm z}}

\newcommand{\vB}{{\bm B}}

\newcommand{\vH}{{\bm H}}

\newcommand{\vJ}{{\bm J}}

\newcommand{\vT}{{\bm T}}

\newcommand{\vY}{{\bm Y}}

\newcommand{\vsigma}{{\bm \sigma}}

\newcommand{\vpsi}{{\bm \psi}}

\newcommand{\vnabla}{{\bm \nabla}}
\newcommand{\vcero}{{\bm 0}}

\newcommand{\hrho}{{\hat{\rho}}}

\newcommand{\esp}[1]{\langle #1 \rangle}

\renewcommand{\Re}{\mathrm{Re}}

\renewcommand{\mfootnote}[1]{\footnote{\rm #1}}

\begin{document}

\title{\textsf{
Positive representations of complex distributions on groups
}}

\author{L. L. Salcedo}
\email{salcedo@ugr.es}

\affiliation{Departamento de F\'{\i}sica At\'omica, Molecular y Nuclear and \\
  Instituto Carlos I de F\'{\i}sica Te\'orica y Computacional, \\ Universidad
  de Granada, E-18071 Granada, Spain.
}


\begin{abstract}
A normalizable complex distribution $P(x)$ on a manifold $\cM$ can be regarded
as a complex weight, thereby allowing to define expectation values of
observables $A(x)$ defined on $\cM$. Straightforward importance sampling,
$x\sim P$, is not available for non positive $P$, leading to the well-known
sign (or phase) problem. A positive representation $\rho(z)$ of $P(x)$ is any
normalizable positive distribution on the complexified manifold $\cM^c$, such
that, $\esp{A(x)}_P = \esp{A(z)}_\rho$ for a dense set of observables, where
$A(z)$ stands for the analytically continued function on $\cM^c$. Such
representations allow to carry out Monte Carlo calculations to obtain
estimates of $\esp{A(x)}_P$, through the sampling $z \sim \rho$.  In the
present work we tackle the problem of constructing positive representations
for complex weights defined on manifolds of compact Lie groups, both abelian
and non abelian, as required in lattice gauge field theories.  Since the
variance of the estimates increase for broad representations, special
attention is put on the question of localization of the support of the
representations.
\end{abstract}


\date{\today}


\maketitle
\flushbottom
\setlength{\unitlength}{1mm}

\tableofcontents

\section{\textsf{Introduction}}
\label{sec:1}

Many a scientific problem, in physics or otherwise, can be reduced to
obtaining the expectation values of observables, assigning a weight to each
existing configuration of some system. When the number of configurations is
large, a Monte Carlo sampling method is often the best option, or even the
only available one in practice \cite{Madras::2002}. However, the route through
importance sampling is blocked when the weights are not definite
positive. This constitutes the well-known sign problem \cite{Troyer:2004ge}.

The sign (or phase) problem arises in many contexts including statistical
mechanics, condensed matter, nuclear physics and quantum field theory, often
related to the presence of fermions in many body systems.  In the context of
lattice gauge field theory the problem arises, for instance in attempting to
study QCD at finite baryonic density. The impediment
is that in the Euclidean formulation the Boltzmann weight is reflection
positive, as required by unitarity \cite{Osterwalder:1977pc}, but not directly
positive in the presence of a chemical potential \cite{Hasenfratz:1983ba}.

Several techniques have been tried to solve or soften the sign problem
\cite{Philipsen:2008zz,deForcrand:2010ys}. Among the potentially exact ones,
one approach is that of reweighting, that is, applying Monte Carlo by sampling
a suitable positive distribution and including the ratio of weights as a
factor in the observable. The method is correct and rigorous but it suffers
from the well-known overlap problem: even for seemingly similar weights,
differences increase exponentially with the size of the system. As a
consequence variances in the estimates increase and the signal-to-noise ratio
becomes negligible \cite{Chandrasekharan:1999cm}.

Another technique aiming at solving the problem exploits the analyticity of
the complex weight in many practical cases, including lattice gauge field
theory. Actually analyticity is routinely used to go from Lorentzian to
Euclidean metrics in those settings. The complex Langevin equation approach
\cite{Parisi:1984cs,Klauder:1983nn} simply applies the stochastic Langevin
equation to the complex case relying on the good analytical properties of the
action, and observables are computed through their analytical extension. This
elegant approach enjoys nice features, above all, that of preserving the
locality of the standard Monte Carlo algorithms, and has been successfully
applied to some practical problems
\cite{Karsch:1985cb,Aarts:2012ft,Sexty:2013ica}. Regrettably, the technique is
not mathematically robust. Even in simple one-degree-of-freedom systems the
algorithm may not converge, or converge to unwanted solutions
\cite{Ambjorn:1986fz,Aarts:2011ax,Salcedo:1993tj,Salcedo:2016kyy}. A recent
review of the present status of the complex Langevin technique can be found in
\cite{Seiler:2017wvd}.

A more recently introduced approach to cope with the sign problem is that of
Lefschetz thimbles \cite{Cristoforetti:2012su,Alexandru:2015sua}. It also
relies on analytic continuation of the action and the observables, using an
optimal deformation of the original real manifold and an additional residual
reweighting. The need of several submanifolds (thimbles), with unknown relative
complex weights, hinders a straightforward application of the method, which is
very promising \cite{Bedaque:2017epw}.

The complex Langevin approach aims at constructing a real and positive
distribution on the complexified manifold, in such a way that the expectation
values of the analytically continued observables correctly reproduce the
expectation values of the original complex weight defined on the real manifold
of configurations of the system. Such a real and positive distribution,
whether originated from complex Langevin or not, was called a {\em
  representation} (of the complex weight) in \cite{Salcedo:1996sa}. 

The explicit construction of direct representations (i.e., constructed without
a complex Langevin approach) was undertaken in \cite{Salcedo:1996sa}. The
existence of positive representations for one-dimensional complex weights was
established in \cite{Weingarten:2002xs}, and for very general complex weights
and manifolds in \cite{Salcedo:2007ji}. Further constructions have been
presented in \cite{Wosiek:2015iwl,Wosiek:2015bqg,Salcedo:2015jxd,%
  Wosiek:2016oah,Seiler:2017vwj,Seiler:2017loi,Ruba:2017agv,%
  Wyrzykowski:2017cfj}.

The two-branch approach in
\cite{Salcedo:2007ji,Salcedo:2015jxd,Seiler:2017vwj,Salcedo:2017tnt,%
  Seiler:2017loi} is particularly suitable in order to obtain localized
representations. This is a major issue in the representation approach since
there is an overlap problem, similar to that of reweighting, related to the
extension of the representation, which reflects on the variance of the Monte
Carlo estimates. Such an approach has been applied in \cite{Salcedo:2015jxd}
to carry out a Monte Carlo sampling with a complex version of the heat bath
method.

Previous works have dealt mainly with complex weights defined on manifolds of
abelian groups, $\R^n$ or $\U(1)^{\times n}$. The case of non abelian groups
is needed in practical applications, such a lattice gauge field theory. This
case was treated in \cite{Salcedo:2007ji} in a rather formal way, showing
existence constructively. In the present work we address the issue of finding
explicit direct representations of complex weights defined on non abelian
matrix groups. The main concepts are revised in Sec. \ref{sec:2}. After a
review of the two-branch approach in $\U(1)^{\times  n}$, we present an
improved prescription to symmetrically treat all the variables, in the
many-dimensional case in Sec. \ref{sec:3}. The case of compact non abelian Lie
groups is considered in Sec. \ref{sec:4}, where formulas are derived for
matrix groups, formally applying to the non compact case too. Obstructions
arise in our approach when some group representations contain singlet
subrepresentations, with respect to the subgroup generated by the element
making the lifting to the complex manifold. This issue is dealt with in
 Sec. \ref{sec:5}, and also some examples are analyzed in detail.
Sec. \ref{sec:6} summarizes our conclusions.

\section{\textsf{Representations of complex probabilities}}
\label{sec:2}

\subsection{\textsf{Definition of representation}}

We consider continuous degrees of freedom throughout.  Let $P(x)$ be a complex
distribution defined on some manifold $\cM$. In applications, $P(x) =
e^{-S(x)}$ where $S(x)$ is the action of the system with configuration $x$.  We
assume that $P$ has a non vanishing normalization, $\int d\mu(x) \, P(x) \not
= 0$.  With some abuse of language, we will refer to $P$ as a {\em complex
  probability}, because expectation values of observables $A(x)$ can be
defined with the same rules as for ordinary (real and positive) probability
densities, i.e.,
\begin{equation}
\esp{A}_P = \frac{\int  P(x) \, A(x) \, d\mu(x) }{\int P(x) \, d\mu(x)  }
,
\end{equation}
where $d\mu(x)$ is a suitable positive measure on $\cM$.

Unfortunately, when $P(x)$ is not positive definite, importance sampling,
$x\sim P(x)$, is meaningless and this prevents the straightforward application
of a Monte Carlo method. This is the well-known sign problem.

Ever since the conception of the complex Langevin algorithm
\cite{Parisi:1984cs,Klauder:1983nn}, one of the approaches devised to sort out
this impediment is to replace the original manifold by its complexified
version $\cM^c$, the observables by their holomorphic extension, $A(z)$, and
the complex probability by an ordinary probability distribution $\rho(z)$
defined on $\cM^c$.\mfootnote{In this work, following
  \cite{Salcedo:1993tj,Salcedo:1996sa,Salcedo:2007ji,Salcedo:2015jxd,%
    Salcedo:2016kyy,Salcedo:2017tnt}, $P$ denotes the complex density defined
  on the real manifold, while $\rho$ denotes the real density defined on the
  complex manifold. The notation exchanging the roles of the symbols $P$ and
  $\rho$ is also frequently used in the literature
  \cite{Okano:1992hp,Aarts:2009uq,Aarts:2011ax, Seiler:2017wvd,
    Seiler:2017vwj}.}

A first obvious condition on $\rho$ is
\begin{equation}
\int_\cM  P(x) \, A(x) \, d\mu(x) =
\int_{\cM^c}  \rho(z) \, A(z) \, d\mu^c(z)
\quad
\text{for all~~} A
.
\label{eq:2.2}
\end{equation}
By definition, a real or complex density $\rho(x)$ fulfilling this condition
will be called a {\em representation} of the complex probability $P$. This
property implies
\begin{equation}
\esp{A(x)}_P = \esp{A(z)}_\rho
,
\label{eq:2.3}
\end{equation}
hence averages obtained from $\rho$ reproduce those of $P$.

An additional condition so that importance sampling can be applied to $\rho$
is to be non negative. A representation $\rho$ will be called a {\em positive
  representation} when $\rho(z)\ge 0$. Therefore we aim at positive
representations of complex probabilities. Although positive representations
are the ultimately interesting ones, we will see that complex representations
also play a role as a mathematical tool.

Regarding \Eq{2.2}, let us remark that the condition can be relaxed by
allowing a different normalization in $P$ and $\rho$.\mfootnote{In
  \cite{Salcedo:1996sa} representations were defined by \Eq{2.3}, while those
  fulfilling also \Eq{2.2} were named {\em unitary} representations.} Also the
requirement ``for all $A$'' in \nec{2.2} really means a suitable (ideally
dense with respect some topology) set of holomorphic test functions, as in
standard distribution theory.  For instance one could take all entire
holomorphic functions, in which case $\rho(z)$ must be of compact support, or
a smaller set such as that of exponentially bounded $A(z)$, allowing more
general $\rho$'s. An even smaller but still practical set of test functions is
that of holomorphic polynomials. In a periodic setting the small set can be
taken as that of finite linear combinations of Fourier modes $e^{ikz}$
($k\in\Z$). For a compact group, the small set of test functions can be taken
as the linear span of the (analytic continuation) of the irreducible
representations of the group.

Finally, let us mention that while $P$ should be normalizable to have
expectation values, complex densities with zero normalization can also be
represented, using the definition in \Eq{2.2}, and they will be useful in the
construction of positive representations of normalized complex densities.

\subsection{\textsf{Existence of positive representations}}

Obviously, complex representations exists for any $P(x)$, for instance
$\rho(z) = P(x)\delta(y)$, where $y$ denotes the coordinates in the imaginary
direction in $\cM^c$. Less trivially, {\em positive} representations also
exist for very general complex probabilities \cite{Salcedo:2007ji}, and the
solution is by no means unique. The non uniqueness follows from the fact that
the set of holomorphic observables constraining $\rho$ is only a subset of all
test functions on the complexified manifold.

An explicit construction for $\cM=\R^n$ has been given in
\cite{Salcedo:2007ji}, as follows. The key observation is that, if
the complex probabilities $P_i(\vx)$, $i\in I$ ($I$ being some index set) admit 
$\rho_i(\vz)$ as positive representations, the complex density
\begin{equation}
P = \sum_i w_i P_i, \qquad w_i \ge 0 
\label{eq:2.4}
\end{equation}
admits
\begin{equation}
\rho = \sum_i w_i \rho_i,
\end{equation}
as a positive representation, provided the sums involved (in presence of
observables) are sufficiently convergent.

To exploit this observation, let us first note that the one-dimensional
complex weight $Q(x)=\delta(x)+\delta'(x)$ admits the following positive
representation on $\C$,
\begin{equation}
q(z) = \frac{1}{8\pi}\left| 1-\frac{z}{2}\right|^2 e^{-{|z|^2/4}}
.
\label{eq:2.6a}
\end{equation}
Clearly $\esp{1}_Q=1$ and $\esp{x}_Q=-1$ and all other $\esp{x^n}$ $n\ge 2$
vanish. On the other hand, reducing $q(z)$ under the group $\U(1)$ (acting as
$z\to \omega z$, ~$|\omega|=1$) it follows that $q(z)$ only contains charges
$0,\pm 1$, hence $\esp{z^n}=0$ for $n\ge 2$. That $\esp{1}_q=1$ and
$\esp{z}_q=-1$ can be checked by direct integration. Hence $\esp{x^n}_Q =
\esp{z^n}_q$ for all $n\in\N_0$ and $q(z)$ is a positive representation of
$Q(x)$.

It can be noted that the representation in \nec{2.6a} is by no means
unique. An easy (but not compulsory) way to comply with the conditions
$\esp{z^n}=0$ for $n\ge 2$ is to take a sufficiently convergent density of the
form $a(|z|)+\Re(b(|z|)z)$ with $a$ real, and the radial functions $a$ and $b$
have a lot of freedom so that the density is non negative and $\esp{1}=1$,
$\esp{z}=-1$ is reproduced. The systematic construction of representations of
the type Gaussian times polynomial for $P(\vx)$ of the same type, or
distributions with support at a single point, in any number of dimensions is
presented in \cite{Salcedo:1996sa}.

Next consider the $n$-dimensional complex density
\be
Q_\vh(\vx)  = \delta(\vx) + \vh \cdot \vnabla \delta(\vx),
\qquad
\vh \in \C^n
,\quad \vx\in\R^n
\,,
\label{eq:2.7}
\ee
which admits the positive representation
\begin{equation}
q_\vh(\vz) = \int_\C  q(\zeta) \, \delta(\vz-\zeta \vh) \, d^2 \zeta 
,\qquad \vz\in\C^n
\,.
\label{eq:2.8}
\end{equation}
The proof of this statement is given in App. \ref{app:A}. This result does
not depend on the concrete choice of $q(z)$ as representation of $Q(x)$, any
other positive representation would do as well. A more localized
representation of $\delta(x)+\delta^\prime(x)$ can be derived using the
two-branch method described below.

The strategy will be to express a generic $P(\vx)$ as a combination of complex
densities of the type $Q_\vh(\vx)$ with positive weights. Without loss of
generality let $P$ be normalized and let $P_0$ be a strictly positive
probability and also normalized,
\begin{equation}
1=  \int_{\R^n} P(\vx) \, d^nx  = \int_{\R^n} P_0(\vx) \, d^nx ,
\quad  P_0(\vx) > 0 
.
\end{equation}
Then $P-P_0$ integrates to zero and can be written as the divergence of a
vector field:
\begin{equation}
P(\vx) = P_0(\vx) + \vnabla \cdot ( P_0(\vx) \vH(\vx))
,
\label{eq:2.6}
\end{equation}
where $\vH(\vx)$ can be chosen in many ways. A particular (non-optimal)
solution can be found by taking $P-P_0=\nabla^2\sigma$, $\sigma$ being the
$n$-dimensional ``Coulomb potential'' created by the ``charge density''
$P-P_0$, and $P_0\vH=\nabla\sigma$ being minus the ``electric field''. The
general solution is found by adding an $n$-dimensional curl to $P_0\vH$.

Clearly \Eq{2.6} can be rewritten in the form \nec{2.4}, namely,
\begin{equation}
P(\vx) =
\int_{\R^n} d^n \vx^\prime \, P_0(\vx^\prime) 
\left(
\delta(\vx - \vx^\prime) + \vH(\vx^\prime) \cdot (\vnabla \delta)(\vx - \vx^\prime)
\right)
,
\label{eq:2.11}
\end{equation}
and this is nothing else than a combination of distributions $Q_\vh$ with
$\vh=\vH(\vx')$ and weight $P_0(\vx^\prime)$. It is straightforward to obtain
a positive representation of $P$ making the replacement $Q_\vh \to q_\vh$ in
\nec{2.11} and using the expression of $q_\vh$ given in \nec{2.8}. In this way
one obtains
\begin{equation}
\rho(\vz) = \int_{\R^n} d^n\vx^\prime \, P_0(\vx^\prime) 
 \int_\C d^2 \zeta\, q(\zeta) \, \delta(\vz-\vx^\prime-\zeta \vH(\vx^\prime))
.
\label{eq:2.12}
\end{equation}
This formula admits a simple interpretation: $\esp{A}_P$ is correctly
reproduced by the average of $A(\vx - \zeta\, \vH(\vx))$, sampling $\vx$ with
$P_0$ and $\zeta$ with $q$.
 
Since $\R^n$ is non-compact, there are technical issues related to convergence
at infinity, they are discussed in \cite{Salcedo:2007ji}.  The analogous
construction for arbitrary compact Lie groups has been given in the same
reference.

\subsection{\textsf{Localization of the support of positive representations}}
\label{sec:2.c}

While the problem of finding positive representations of generic complex
distributions is formally solved, the impediments for systems of large
dimensionality remain in practice.  Indeed, the vector field $\vH$ is not easy
to obtain in an usable form. Even more importantly, in general, the magnitude
of $\vH$ will scale as $|P|$ as the number of degrees of freedom (or volume)
increases. Since the action $S$ scales as the volume, this implies an
exponential growth in $\vH$ which in turn entails an exponential growth in the
size of the support of the representation $\rho$ and so in the dispersion of
the random variable $\vz$ in $A(\vz)$. This would translate into an
exponentially large variance in the Monte Carlo estimates.

This is an important aspect of the representation approach: in the standard
case of positive probabilities, the sampling $x \sim P$ is uniquely defined by
$P$.\mfootnote{The influence of the concrete observable $A$ on the sampling,
  in order to reduce the variance, is of academic interest only, first because
  sampling is expensive and many observables are to be considered, and second
  because $P=e^{-S}$ behaves exponentially with respect to typical observables
  (including $S$) and so sampling $P$ is mandatory.} This is no longer true
when the estimate is obtained by means of a representation since many
different representations exist. These are all formally equivalent (as all of
them fulfill \Eq{2.2}) but they can be very different regarding the variance
of the estimates obtained from them.\mfootnote{The test functions involved in
  computing the variance are not holomorphic, so their expectation values are
  not protected by the equality \nec{2.2} and depend on the concrete
  representation.} Ideally one would like a $\rho$ with a support as localized
as possible in order to reduce the dispersion. This problem is analogous to
that in the reweighting approach, where a maximum overlap is desirable. A
complete overlap is not possible if $P$ is complex, and also in the
representation approach a perfect localization of $\rho$ on the real manifold
is not attainable.

Since observables tend to grow wildly as one departs from the real manifold,
representations close to it are preferable in general. The {\em width} of a
representation $\rho$ can be defined as the size of its support in the
imaginary direction, and for a given complex probability there are bounds on
how narrow any positive representation of it can be. As one would expect, the
more complex (in the sense of less positive definite) a complex probability is
the wider is its narrowest positive representation. Not surprisingly,
obtaining wider (and so worse quality) representations poses no
problem.\mfootnote{Applying an isotropic diffusion process to any positive
  representation produces another, less localized, positive representation of
  the same complex probability \cite{Salcedo:1996sa}.}

Regarding the localization of the support of any positive representation of a
given complex probability, a general observation can be made
\cite{Salcedo:2015jxd}: for any observable $A$, the support of $\rho$ must
contain values of $|A(z)|$ larger than $|\esp{A}_P|$ (note that this quantity
is independent of the choice of $\rho$).\mfootnote{This simple consideration,
  for instance, rules out that the complex Langevin algorithm could produce a
  proper representation for the action $S(x) = x^4/8+2ix$
  \cite{Salcedo:2015jxd}.} 

In particular a concrete bound follows (in the one-dimensional case but can be
extended to higher dimensions). Let us assume that the support of $\rho(z)$ is
entirely contained in a horizontal strip $Y_2 < y < Y_1$. Then
$|e^{-ikz}|=e^{ky} \le e^{kY_1}$ for $k>0$, implies $|\esp{e^{-ikz}}_\rho| \le
e^{kY_1}$. Because $\esp{e^{-ikz}}_\rho = \esp{e^{-ikx}}_P$ does not depend
on the representation, this inequality $\forall k>0$ puts a constraint on the
admissible values of $Y_1$. An analogous consideration $\forall k<0$ and
$Y_2$, leads to the following bounds on the support of any positive
representation
\begin{equation}
Y_1 \ge \max_{k>0}\left(\frac{1}{k}\log|\tilde{P}(k)|\right)
,
\quad
Y_2 \le  \min_{k<0}\left(\frac{1}{k}\log|\tilde{P}(k)|\right)
,
\label{eq:2.10}
\end{equation}
where $\tilde{P}(k)=\esp{e^{-ikx}}_P$ is the Fourier transform of $P(x)$. In
practice, these bounds can be quite tight for typical $P$'s
\cite{Salcedo:2015jxd}.

With some ingenuity additional conditions can be imposed on the support
of a positive measure $\rho$ representing a complex probability $P$. For
instance, for any observable $A(x)$, let $a \equiv \esp{A}_P$, and let two
nonempty complementary regions in $\cM^c$ be defined by $\mathcal{A}_> = \{
\Re (A(z)) \ge \Re(a) \}$ and $\mathcal{A}_< = \{ \Re (A(z)) < \Re(a) \}$ (we
exclude the trivial case of a constant $A$).  Then the relation
\begin{equation}
\esp{\Re ( A ) }_\rho = \Re (a)
\label{eq:2.4a}
\end{equation}
requires that the support of $\rho$ must have some overlap with both regions
as it cannot be {\em entirely} contained in any of them. The fulfillment of
this condition for all observables $A$ puts constraints on the allowed support
of positive (or more generally real) representations. Of course, taking
$e^{-i\theta}A$, the same consideration holds for $\Re_\theta(A) \equiv
\Re(e^{-i\theta} A )$~($\theta\in\R$), and for $\Im(A)$ in particular.

The usefulness of this kind of relations can be seen in the following example.
Let $P(x)= e^{-s(x)}$ with $S(x)=x^4-2x^2-2ix$. For this complex probability
$\esp{x}=-7.83 i$. Since this value is below the real axis, any positive
$\rho(z)$ representing $P(x)$ must have some support below the real
axis. However, if one applies a standard complex Langevin prescription, the
stationary solution for $\rho$ will be above the real axis: the velocity drift
points upwards along the real axis so the complex Langevin walker can never
cross the real axis once she is above it. This localization argument exposes
the failure of complex Langevin in this case without an explicit simulation of
the stochastic process.

Summarizing, positive representation exists for arbitrary or very general
complex probabilities, and localized representations are highly preferable
from the point of view of Monte Carlo calculations. It
is also noteworthy that one can impose on the representations the same
symmetries enjoyed by the complex probability itself provided the
symmetrization procedure is compatible with the analytic extension, which is
often, if not always, the case. This property will be exploited in the
construction of representations, namely, by decomposing the complex probability
defined on a group as a sum of (often irreducible) group representations.

\section{\textsf{Localized representations of abelian groups}}
\label{sec:3}

The complex probabilities considered in this section are defined on $\R^n$ or
periodic versions of it, so they can be viewed as complex probabilities on
abelian groups, namely, $(\R^n,+)$ or $\U(1)^{\times  n}$ or mixed cases of
them.

We first review the construction of localized representations carried out in
\cite{Salcedo:2015jxd}. A similar construction has been derived independently
by Seiler and Wosiek in \cite{Seiler:2017vwj}. The one-dimensional and higher
dimensional cases are discussed. Subsequently, a more systematic and
satisfactory treatment of the higher dimensional case is introduced.

An important feature of the representations discussed here is that their
support is composed of (a finite number of) parallel copies of the real
manifold, at different heights in the imaginary direction. Therefore, these
representations can be used with any holomorphic observable, regardless of how
wildly such observable may behave in the deep imaginary region. Analogous
constructions will be obtained for complex measures defined on more general
groups in the next section.

\subsection{\textsf{Two-branch representations in one-dimension}}
\label{sec:3.1}

Consider $P(x)$ defined on $\U(1)$. The case $x\in\R$ is completely analogous
in most respects and is described in  \cite{Salcedo:2015jxd}. We use the
normalization
\begin{equation}
1 = \int_0^{2\pi} \frac{dx}{2\pi} \, P(x)
\end{equation}
and assume $P$ to be normalized throughout the construction.

A suitable set of holomorphic test functions is $e^{-ikx}$, hence we aim at
finding a positive representation $\rho(z)$ such that ($z=x+iy$)
\begin{equation}
 \int_0^{2\pi} \frac{dx}{2\pi} \, P(x) \, e^{-ikx}
=
 \int_0^{2\pi}  \frac{dx}{2\pi} \int_{-\infty}^{+\infty} dy\, \rho(z) \,
 e^{-ikz}
,\quad
\forall k \in  \Z
.
\label{eq:3.2}
\end{equation}

As said, there are many solutions for $\rho$ and we favor the most localized
ones. A sensible support is a strip parallel to the real axis, because a
finite estimate would result even for holomorphic test functions with a wild
behavior in the deep imaginary region. Even better one can choose the support
to be lines parallel to the real axis. Clearly, a single line would not be
sufficient for generic complex densities $P(x)$, however, it turns out that
two lines are sufficient. This makes sense because two real functions (one
function on each line) can carry the same information as a single complex one,
$P(x)$.

The (symmetric) {\em two-branch} representation is of the form
\begin{equation}
 \rho(z) = 
Q_+(x) \delta(y - Y) + Q_-(x) \delta(y + Y) , \quad Y>0
\label{eq:3.3}
\end{equation}
where $Q_\pm(x)$ are two real and positive periodic functions (or
distributions). That is, $\rho$ has support on the two horizontal lines $y=\pm
Y$, parallel to the real axis. Each of the two branches is a copy of the real
manifold. The width is $2Y$ and this is a parameter to be chosen in the
construction.

The workings of the two-branch representation can be seen by multiplying both
sides of \nec{3.3} by a generic holomorphic test function. Upon integration
\begin{equation}
\int_0^{2\pi} \frac{dx}{2\pi} \int_{-\infty}^{+\infty} \rho(z) A(z) =
\sum_{\sigma=\pm} \int_0^{2\pi} \frac{dx}{2\pi} Q_\sigma (x) A(x + i\sigma Y)
.
\label{eq:3.4a}
\end{equation}
Introducing the normalizations of $Q_\pm(x)$
\begin{equation}
N_\pm \equiv \int \frac{dx}{2\pi} Q_\pm(x)
,
\end{equation}
and using the representation property $\esp{A(x)}_P = \esp{A(z)}_\rho$,
\nec{3.4a} becomes
\begin{equation}
\esp{A(x)}_P
=
\sum_{\sigma=\pm} N_\sigma \esp{A(x+i\sigma Y)}_{Q_\sigma}
.
\label{eq:3.6c}
\end{equation}
The interpretation of this equation is that $\esp{A(x)}_P$ can be obtained
from the averages of $A(x \pm i Y)$ with $x\sim Q_\pm$. 

For given $Y$, the functions $Q_\pm$ must be chosen to comply with
\nec{3.2}. In fact the two functions $Q_\pm(x)$ are (almost) uniquely
determined by the requirement of them being {\em real} (for real $x$). To see
this let us introduce the Fourier modes
\begin{equation}
P(x) = \sum_k \tilde{P}_k \, e^{ikx}
,
\qquad
Q_\pm(x) = \sum_k \tilde{Q}_{\pm,k} \, e^{ikx}
.
\end{equation}
Use of \Eq{3.3} in \nec{3.2} yields the equations
\begin{equation}
\tilde{P}_k = e^{kY}\tilde{Q}_{+,k} + e^{-kY}\tilde{Q}_{-,k}
\quad \forall k\in\Z\,.
\label{eq:3.5}
\end{equation}
The reality conditions on $Q_\pm(x)$ imply $\tilde{Q}_{\pm,k}^* =
\tilde{Q}_{\pm,-k}$ and allow to write a second set of equations
\begin{equation}
\tilde{P}^*_{-k} = e^{-kY}\tilde{Q}_{+,k} + e^{kY}\tilde{Q}_{-,k}
\quad \forall k\in\Z\,.
\label{eq:3.6b}
\end{equation}
The two sets yield the solution
\begin{equation}
\tilde{Q}_{\pm,k}
=
\pm \frac{ e^{\pm k Y} \tilde{P}_k -  e^{\mp k Y} \tilde{P}_{-k}^* }{2 \sinh(2kY)}
\quad (k\not=0)
.
\label{eq:3.6}
\end{equation}

The solution is unique except for $k=0$ which is not determined. Indeed the
rhs of the two equations \nec{3.5} and \nec{3.6b} are identical for $k=0$ and
the system is compatible owing to the fact that $P(x)$ has a real normalization
($\tilde{P}_0=1$) in such a way that the two lhs also coincide.  A similar
situation will be found in the treatment of higher dimensions and of non
abelian groups by means of two-branch representations, not only for the
constant mode, but also for other non trivial modes. In general the equations
obtained will be compatible only for appropriate choices of the support of
$\rho$. This problem is discussed later in this section for the higher
dimensional case and in Sec. \ref{sec:5} for non abelian groups.

The {\em zero mode} components $\tilde{Q}_{\pm,0}$ are the normalizations of
the two functions, $N_\pm$ and can take any values subject to the conditions
\begin{equation}
N_++N_-=1
,\qquad N_\pm \ge 0
.
\end{equation}

From the Fourier components, it follows that the functions $Q_\pm(x)$ have an
improved behavior, as compared to $P(x)$, as regards to smoothness. This
comes about from the extra factor $e^{-|k|Y}$ in $\tilde{Q}_{\pm,k}$ with
respect to $\tilde{P}_k$, for large $|k|$. In particular, if $P(x)$ happens to
be analytic on $\R$, say within a strip of width $Y_1>0$, $Q_\pm(x)$ are
analytic within a strip of width $Y+Y_1$. That is, the functions taking values
$Q_\pm(x)$ on the lines $z=x \pm i Y$, can be analytically extended to a
region containing the real axis. This allows to write the important relation
\begin{equation}
P(x) = Q_+(x - i Y) + Q_-(x + i Y)
\,.
\label{eq:3.4}
\end{equation}
This can be shown as follows: \nec{3.6c} states that
\begin{equation}
\int_0^{2\pi} \frac{dx}{2\pi} P(x) A(x) 
=
\sum_{\sigma=\pm} \int_0^{2\pi} \frac{dx}{2\pi} Q_\sigma(x) A(x+i\sigma Y)
.
\end{equation}
$Q_\pm(x)$ admitting an analytic extension from $\R$ to $\R \mp i Y$, allows
to shift the variable $x$ in the integral to write
\begin{equation}
\int_0^{2\pi} \frac{dx}{2\pi} P(x) A(x) =
\sum_{\sigma=\pm} \int_0^{2\pi} \frac{dx}{2\pi} Q_\sigma(x-i\sigma Y) A(x)
.
\end{equation}
Since this holds for any test function \Eq{3.4} follows.

In turn \Eq{3.4} leads to \nec{3.6c}, as is easily shown.  Therefore \Eq{3.4}
contains the information that $\rho(z)$ is a representation and specifically
one of the two-branch type. The analysis in terms of Fourier modes shows that
the solution of \nec{3.4} (plus the reality conditions) is essentially
unique. The ambiguity in the constant modes $N_\pm$ is seen in \nec{3.4} as
the freedom to add a constant function to $Q_+$ and subtract it from $Q_-$,
without violating the equation.

The relation \nec{3.4} follows immediately from using the two-branch form
\nec{3.3} in \Eq{3.30} of Sec.  \ref{sec:III.D.1}.  The formulation based on
\nec{3.4} will be preferable in the higher dimensional abelian and non abelian
cases, as it avoids the need to discuss Dirac deltas on the manifold of the
complexified group, instead only copies of the original group manifold are
required. It is true that \nec{3.4} assumes analyticity of $P(x)$ on the real
manifold, but this is hardly a restriction: one can treat $P(x)$ as the limit
of a truncated sum of Fourier modes, and the relations derived for finite Fourier
modes, like those in \nec{3.5}, will be preserved as the cutoff is removed, and
the same argument will apply for other groups, in which $P$ is decomposed
into irreducible representations of the group.

To obtain a positive two-branch representation we still have to show that the
$Q_\pm(x)$ are non negative choosing $Y$ appropriately. By construction
$Q_\pm(x)$ are real for any value of $Y$. In general they are not positive
definite and diverge for small $Y$, except when $P(x)$ is real. In that case
\begin{equation}
\tilde{Q}_{\pm,k} = \frac{\tilde{P}_k }{2 \cosh(kY)}
\qquad\text{(real $P(x)$)}
\end{equation}
and $Q_\pm(x)\to \frac{1}{2}P(x)$ as $Y\to 0$.

Going in the opposite direction of increasing $Y$, we have already noted the
presence in \nec{3.6} of the factor $e^{-|k|Y}$, as there are two powers of
$e^{|k|Y}$ in the denominator and only one in the numerator. This implies that
as $Y$ increases the modes $\tilde{Q}_{\pm,k}$ will be quenched, provided only
that $\tilde{P}_k$ is exponentially bounded, that is, if $|\tilde{P}_k|< K
e^{Y_1|k|}$ for some $K,Y_1>0$. This is an extremely lax condition which
includes the ordinary distributions. For {\em sufficiently large} $Y$, all non
zero Fourier modes in \Eq{3.6} become arbitrarily small hence, taking
$N_\pm > 0$, it follows that eventually $N_\pm$ dominate the Fourier sum
and $Q_\pm(x)$ are guaranteed to be positive. This shows that essentially any
periodic complex probability admits a positive representation of the
two-branch type.  Explicit examples of representations of the two-branch type
can be found in \cite{Salcedo:2015jxd}.

As already noted, in practice it is advantageous to have a width as small as
possible. The prescription to achieve this is the following:\mfootnote{This an
  improvement over \cite{Salcedo:2015jxd}, where it was not realized that
  $q_\pm$ are necessarily non positive, since $Q_\pm(x)-N_\pm$ has zero
  normalization.} starting from the bounds in \Eq{2.10}, $Y$ can be
continuously increased. Eventually, for some critical value $Y=Y_c$
\begin{equation}\begin{split}
&
q_++q_- =  -1
\,,
\\ &
q_\pm \equiv \min_x \sum_{k\not=0} \tilde{Q}_{\pm,k} \, e^{ikx}
= \min_x (Q_\pm(x)-N_\pm)
.
\label{eq:3.10a}
\end{split}
\end{equation}
For $Y \ge Y_c$, suitable $0 \le N_\pm \le 1$ exist so that $Q_\pm(x)$ are
positive for all $x$. In particular for $Y=Y_c$, $\min_x Q_\pm(x) = 0$.

The construction in $\R$ (as opposed to $[0,2\pi]$) is quite similar, the main
difference being that the freedom in sharing zero modes between the two sheets
$y=\pm Y$ no longer exists \cite{Salcedo:2015jxd}. We discuss further the
noncompact case at the end of Sec. \ref{sec:III.D.2}.

It can be noted that we have chosen as support of our representation exactly
two horizontal lines and equidistant from the real axis, $y=\pm Y$.  As
discussed in \cite{Salcedo:2015jxd} an asymmetric choice is possible but in
practice no substantial gain is achieved by doing that (for generic complex
probabilities). So we favor simplicity in our construction in order to
facilitate its extension to more complicated scenarios. Incidentally, the use
of two more general curves as branches, not necessarily horizontal lines, is
also possible, and this can be used in principle to avoid certain regions,
e.g., allowing to treat test functions with singularities at prescribed
points. However the treatment is considerably more complicated as the
zero-mode ambiguity is no longer an additive constant to be applied to the
weights $Q_\pm$.

Another question is the use of more branches, $y=Y_1,\ldots, Y_n$.  Also
nothing is gained in practice. Moreover, since one must impose positivity on
each branch separately, this implies a larger number of conditions which
translate into larger values of $Y_i$ (and so larger variances). In
\cite{Salcedo:2007ji} each Fourier mode $a_k e^{ikx}$ was treated
separately. This is legitimate but not optimal. Since a single Fourier mode
has zero normalization (except $k=0$) one must share the total normalization
of $P$ (namely, $1$) among the Fourier modes, and obtain a positive
representation of each $n_k + a_k e^{ikx} $. For a fixed amplitude $a_k$, the
smaller the normalization $n_k$, the wider the representation (larger $Y$). So
the sharing among modes, $1=\sum_kn_k$, must be optimized and even so,
imposing positivity for the representation of each separate mode requires
larger values of $Y$. The great advantage of the two-branch approach of
\cite{Salcedo:2015jxd} is that all the modes are added on the same branch
(same support) and they compensate each other to have a positive function
requiring a minimal common width.

\subsection{\textsf{Two-branch representations in higher dimensions}}
\label{sec:3.2}

The above construction can be generalized to functions defined on the torus
$[0,2\pi]^n$, or equivalently $\U(1)^{\times  n}$, although this is not completely
straightforward.

\subsubsection{\textsf{Strict two-branch approach}}
\label{sec:3.2.1}

For normalized $P$, one can tentatively propose
\begin{equation}
\rho(\vz) = 
Q_+(\vx) \delta(\vy - \vY) + Q_-(\vx) \delta(\vy + \vY)
,
\label{eq:3.10}
\end{equation}
where the two functions $Q_\pm(\vx)$ are positive and the construction depends
on the parameters $\vY=(Y_1,\ldots,Y_n)$. The representation condition is
equivalent to requiring
\begin{equation}
P(\vx) =
Q_+(\vx - i \vY) + Q_-(\vx + i \vY)
,
\label{eq:3.11}
\end{equation}
and in terms of the Fourier modes this implies (demanding that $Q_\pm(\vx)$
should be real)
\begin{equation}
\tilde Q_{\pm,\vk}
=
\pm \frac{ e^{\pm \vk\cdot\vY} \tilde{P}_\vk - e^{\mp \vk\cdot\vY} \tilde{P}_{-\vk}^* 
}{
2 \sinh(2\vk\cdot\vY)}
\qquad (\vk\cdot\vY\not=0)
.
\label{eq:3.6a}
\end{equation}
Note that $\tilde Q_{-,\vk}$ is just $\tilde Q_{+,\vk}$ with $-\vY$ instead of
$\vY$.

Once again, the constant modes,
\begin{equation}
N_\pm = \int \frac{d^nx}{(2\pi)^n} Q_\pm(\vx)
,
\end{equation}
are not fixed since, being constant under analytic extension, they can be
moved freely between the two branches in \Eq{3.11}. Also, for large enough $\vY$
(assuming $\vk\cdot\vY\not=0$) all non constant Fourier modes become small and
the distributions $Q_\pm(\vx)$ eventually become positive for positive
$N_\pm$.

Clearly the singular modes, i.e., those with $\vk\cdot\vY = 0$, pose a
problem. This is for the same reason $\vk=0$ is special: since $Q_\pm$ are
real, if one integrates over $\vx$ on both sides of \Eq{3.11} the resulting
equation is only consistent if the normalization of $P$ is also
real. Equivalently, the zero (constant) mode is unchanged by the shifts $ \vx
\to \vx \pm i \vY $ from the real to the complex manifold. By the same token,
the singular modes with $\vk\cdot\vY=0$ are not affected by the complex shift
and the equation is only consistent if $P(\vx)$ happens to be real for those
particular modes. For the zero mode, the reality condition is fulfilled due to
our previous requirement that $P$ should be normalized, but no analogous
property exists fixing the remaining singular modes.

An easy solution would be to take for the components of $\vY$ suitable
irrational numbers in such a way that the combination $\sum_i k_iY_i$ can
never be exactly zero (e.g. $\vY=(1,\sqrt{2},\sqrt{3})$). However, such
prescription is rather arbitrary and has several drawbacks: i) although
$\vk\cdot\vY$ would not be exactly zero it could be arbitrarily small when
many modes are relevant and this is numerically problematic.\mfootnote{In fact,
  in the noncompact case ($\cM=\R^n$ rather than a torus) $\vk$ is continuous
  and $\vk\cdot\vY = 0$ would not be avoided.} ii) As the problem worsens when
all components of $\vY$ are similar, this suggests using very dissimilar
components.  Unfortunately, positiveness of $Q_\pm(\vx)$ requires a
sufficiently large vector $\vY$ but too large values entail large variances;
dissimilar values of the components of $\vY$ imply that some of these
components would be larger than necessary (to allow the shorter components to
be sufficiently large). iii) Most importantly, if the various degrees of
freedom represented by the variables $\vx=(x^1,x^2,\ldots,x^n)$ play a similar
role in the action (a similarity that is often enforced by concrete symmetries
of the action) one would request that $\vY$ should also contain similar
components for all of them, without ad hoc variation from one component to
another, with no basis on the action or the physical problem at hand.

\subsubsection{\textsf{Uniform two-branch approach}}
\label{sec:3.2.2}

A better solution is to use different displacement vectors $\vY$ for different
Fourier modes.\mfootnote{Such possibility is noted in \cite{Seiler:2017vwj} and
  it was also present in \cite{Salcedo:2007ji} where each Fourier mode is
  treated independently.}  Implicitly this implies to introduce further
branches, i.e., further copies of the real manifold. In order to encompass the
uniformity criterion noted above, in which all variables should play a similar
role, a natural prescription is to introduce $2^n$ branches, a duplication for
each degree of freedom. Each branch is characterized by a vector of $n$ bits,
$\vsigma=(\pm,\ldots,\pm)$, so that
\begin{equation}
\vY = (\pm Y,\ldots, \pm Y) = Y \vsigma
.
\end{equation}
Correspondingly, there are $2^n$ real and positive functions $Q_\vsigma(\vx)$
defined on the real manifold, and the representation condition becomes
\begin{equation}
P(\vx) = \sum_{\vsigma= (\pm,\ldots,\pm)} Q_\vsigma (\vx - i \vsigma Y)
.
\end{equation}
Effectively, the full configuration on the complexified manifold is described
by a real and positive function $Q({x_1,\sigma_1},\ldots,{x_n,\sigma_n})$.
Each degree of freedom is augmented with an additional bit.\mfootnote{In {\em
    counting} degrees of freedom, this would be equivalent to duplicating the
  original coordinate range by joining two copies of it, for each
  coordinate. For instance, $[0,2\pi] \to [0,4\pi]$, or $\R^+ \to
  \R$. Unfortunately this picture does not work topologically, as the copies,
  say $[0,2\pi]$ and $[2\pi,4\pi]$, would not be related through any
  continuity condition.}

Our proposal is to share each Fourier mode $\vk$ among $2^{m+1}$ branches,
where the value of $m$ and the concrete branches depend on the mode. For any
such branch $\vsigma$, $Q_\vsigma$ is given by $Q_+$ in \Eq{3.6a} with $\vY =
Y \vsigma$ and an additional factor $1/2^m$. The concrete assignation of
branches is as follows.

$a)$ For a Fourier mode $\vk = (k_1,\ldots,k_n)$ with {\em all $k_i$ different
  from zero}, only two branches are involved ($m=0$) and \Eq{3.6a}
applies. One of the branches is that with $\sigma_i = \sign(k_i)$,
or equivalently, $k_i Y_i > 0$\, for each $i$. The other branch is the
opposite one, with all $k_i Y_i < 0$. This assignation of branches certainly
guarantees that $\vk\cdot\vY$ is never zero and complies with the uniformity
criterion.

$b)$ For Fourier modes in which some (but not all) of the $k_i$ are zero: for
the subset of $k_i$ which are not zero the rule for the assignation of branch
is as above (i.e., all $\sigma_i=\sign(k_i)$ or all
$\sigma_i=-\sign(k_i)$). For the vanishing $k_i$ there is an ambiguity
(completely analogous to the ambiguity in the choice of $N_\pm$). The most
symmetric prescription is to assign half of the strength to each of the two
possibilities $\sigma_i=\pm 1$. So a Fourier mode in which $k_i$ vanishes for
$m$ values of $i$ will be distributed among $2^{m+1}$
branches. Correspondingly $Q_+$ in \Eq{3.6a} picks up a factor $1/2^m$.

$c)$ The constant mode, $\vk=0$, is equally distributed among the $2^n$
branches, that is $N_\vsigma = 1/2^n$, where $N_\vsigma$ is the normalization
of $Q_\vsigma$.\mfootnote{Any other distribution with
  non negative $N_\vsigma$ would be valid, perhaps allowing a smaller
  $\vY$. The one proposed here is just the simplest one, and this also true
  for the prescription adopted in the case $b)$.}

Equivalently, for all $\vsigma$ and $\vk$, $Q_+$ in \Eq{3.6a} applies (with
$\vY=Y\vsigma$) but with an additional factor. The factor is $1/2^m$ if $m$
values $\sigma_i k_i$ vanish while the other are all positive or all
negative. Otherwise the factor is zero.

\begin{figure}[t]
\begin{center}
\includegraphics[height=50mm]{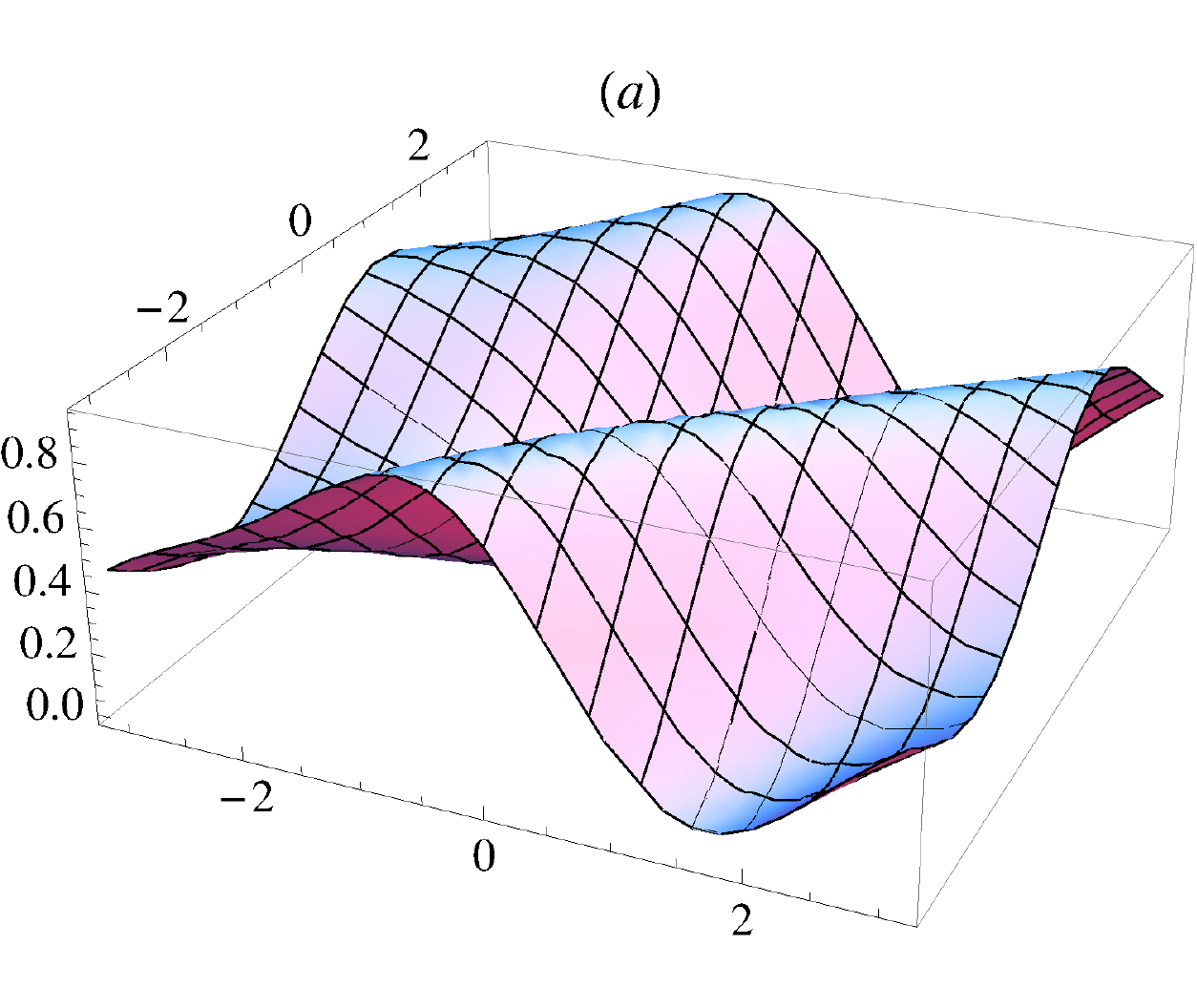}
\includegraphics[height=50mm]{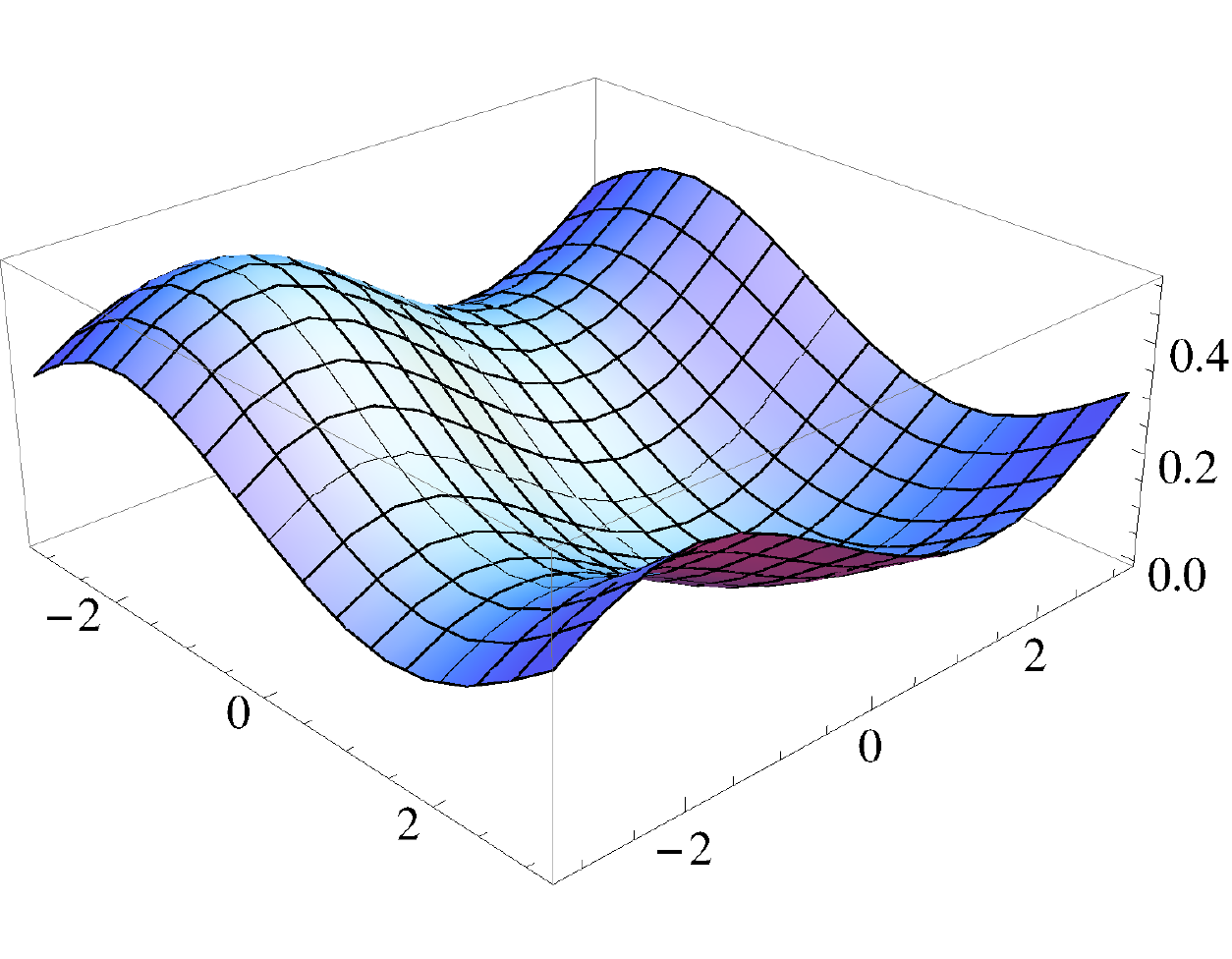}
\end{center}
\caption{Representations of $(1+\beta \cos(x_1))(1+\beta \cos(x_2))(1+\beta
  \cos(x_1-x_2))$ for $\beta=i$.  ~$(a)$ Exactly two sheets with
  $Y_2/Y_1=\sqrt{2}$ and $Y_2=3.87$. The function $Q_+(\vx)$ is displayed.  
  ~$(b)$ Four sheets with $Y_1,Y_2 = \pm Y$ and $Y=1.51$. The function
  $Q_{++}(\vx)$ is displayed.  }
\label{fig:1}
\end{figure}

The $Q_\vsigma$ will be non negative for $Y>Y_c$, with $Y_c$ obtained from the
condition
\begin{equation}
\sum_{\vsigma}q_\vsigma = -1,
\qquad
q_\vsigma \equiv \min_{\vx} (Q_\vsigma(\vx) - N_\vsigma)
\,.
\end{equation}

As illustration, consider the two-dimensional distribution
\begin{equation}
P(x_1,x_2) \propto
 (1+\beta \cos(x_1))(1+\beta \cos(x_2))(1+\beta \cos(x_1-x_2))
,
\end{equation}
with $\beta=i$. This distribution admits a positive representation using
exactly two sheets with an asymmetric choice $\vY \propto (1,\sqrt{2})$. The
relation $Q_-(\vx) = Q_+(-\vx)$ holds automatically. The optimal width, that
is, such that $\min_{\vx} Q_+(\vx)=0$, is obtained as $\,\vY = (2.74,
3.87)$. The branch $Q_+(\vx)$ is displayed in Fig.~\ref{fig:1}$a$.

The alternative construction with four sheets, $Q_{\pm\pm}(\vx)$, attains a
positive representation with $\,\vY = (\pm 1.51, \pm 1.51 )$, which having a
smaller width represents an improvement over the previous asymmetric
construction. Symmetry under $(x_1,x_2) \to (x_2,x_1)$ is automatic, and also
$Q_{-\mp}(\vx) = Q_{+\pm}(-\vx)$ is fulfilled.  The branch
$Q_{++}(\vx)$ is displayed in Fig.~\ref{fig:1}$b$, the branch $Q_{+-}(\vx)$
has a similar shape, up to a reflection.

\subsection{\textsf{Representations from convolutions}}

The representations just described can be written as convolutions.  Let us
consider first the simple case in which problems coming from $ \vk \cdot
\vY=0$ can be neglected. The zero mode is treated separately as this singular
term is always present. Straightforward reconstruction of the Fourier sum
using the components in \Eq{3.6a} gives
\begin{equation}
Q_+(\vx) = 
N_+ +
2\Re \sum_{\vk\not=0}
 \frac{ e^{i\vk\cdot\vx} e^{\vk\cdot\vY} 
}{
2 \sinh(2\vk\cdot\vY)}
\int \frac{ d^nx^\prime }{ (2\pi)^n } e^{-i\vk\cdot\vx^\prime} P (\vx^\prime )
.
\label{eq:3.17}
\end{equation}

In order to proceed, let us introduce the following function
\begin{equation}
\chi (\Omega) \equiv \frac{ \Omega }{ \Omega^2-\Omega^{-2} }
\label{eq:3.18}
\end{equation}
and also the distribution
\begin{equation}
C(\vx;\vY) \equiv 
\sum_{\vk \not = 0}  e^{i\vk\cdot\vx} \chi(e^{\vk\cdot\vY})
.
\label{eq:3.19}
\end{equation}
This allows to express $Q_\pm(\vx)$ as convolutions:
\begin{equation}\begin{split}
Q_\pm(\vx) &= 
N_\pm +  2 \Re ( C(\vx; \pm\vY) * P(\vx) )
\\
&= 
N_\pm +  2 \Re ( C( \pm \vx; \vY) * P(\vx) )
.
\label{eq:3.20}
\end{split}\end{equation}
(For convenience we denote $f(x)*g(x)$ what is usually denoted $(f*g)(x)$.)
As is readily verified, the identities
\begin{equation}\begin{split}
& \chi(\Omega)\Omega + \chi(\Omega^{-1})\Omega^{-1}  = 1
,\\
& \chi(\Omega)\Omega^{-1} + \chi(\Omega^{-1})\Omega  = 0
,
\end{split}\end{equation}
guarantee the fulfillment of \Eq{3.11}. It should be noted that the expression
using {\em real part} in \Eq{3.20} refers only to {\em real} $\vx$. Of course
the analytic extension implied in \Eq{3.11} has to be applied {\em after}
the real part is expanded in \Eq{3.20} as a linear combination of $P(\vx)$ and
$P^*(\vx)$.

We can turn now to the improved construction using $2^n$ branches. 
Again the zero mode is treated separately, only subject to the conditions
\begin{equation}
1 = \sum_\vsigma N_\vsigma 
,\qquad
N_\vsigma \ge 0
.
\label{eq:3.8a}
\end{equation}
For the remaining Fourier modes the expression in \Eq{3.17} still holds with
$\vY=Y\vsigma$ and taking into account that not all modes contribute to each
branch $\vsigma$: In principle, a given mode $\vk=(k_1,\ldots,k_n)$
contributes only to the branch with all $\sigma_i$ equal to $\sign(k_i)$ or
all opposite. When some $k_i$ are zero, these are equally distributed between
the $\sigma_i=1$ and $\sigma_i=-1$ options.

In this way, the functions $Q_\vsigma(\vx)$ can be written as convolutions in
the form
\begin{equation}
Q_\vsigma(\vx) = 
N_\vsigma +  2 \Re ( C(\vsigma*\vx; Y) * P(\vx) )
,
\quad
\vsigma =(\pm,\ldots,\pm)
,
\label{eq:3.20a}
\end{equation}
where we have defined
\begin{equation}
\vsigma*\vx \equiv (\sigma_1 x^1,\ldots, \sigma_n x^n)
,
\end{equation}
and
\begin{equation}
C(\vx;Y) \equiv 
\sum_\vk  e^{i \vk \cdot \vx} \chi(e^{ Y \sum_i k_i })
\,\Theta(\vk)
.
\end{equation}
The function $\Theta(\vk)$ selects the Fourier modes contributing to the
branch $(+,\ldots,+)$,
\begin{equation}
\Theta(\vk) \equiv
\left\{\begin{matrix}
\prod_{i=1}^n \theta(k_i) +
\prod_{i=1}^n \theta(-k_i)
& ( \vk\not = 0 )
\\
0 & ( \vk = 0 )
\end{matrix}\right.
,
\end{equation}
$\theta(x)$ being the Heaviside step function with $\theta(0)=1/2$. The
function $\Theta(\vsigma*\vk)$ does the same job for a branch $\vsigma$.

\subsection{\textsf{Complex representations and linearity}}
\label{sec:III.D}

\subsubsection{The projection operator $K$}
\label{sec:III.D.1}

Loosely speaking, a (in general complex) distribution $\rho$ on the
complexified manifold defines, through \Eq{2.2}, an associated complex
probability $P$ on the real manifold. Let us denote by $K$ the corresponding
projection operator, that is,
\begin{equation}
K \rho = P
.
\end{equation}
Of course, as for the observables, this assumes some class of sufficiently
well behaved $\rho$. 

\newcommand{\mS}{\mathrm{S}}
To make precise definitions, let us consider a periodic setting in one
dimension, hence the real manifold is the circle $\mS^1\cong [0,2\pi]$ and the
complex manifold is the cylinder $\mS^1\times\R$. As space of test functions
on the cylinder, $A(z)$, let us take the linear span of the Fourier modes
$e^{ikz}$, this space will be denoted $\cD_c$.  The space of densities
$\rho(z)$ can be chosen in many ways. A sufficiently general space is that of
Schwartz distributions on the cylinder and with bounded support in it. Let us
denote this space $\cR$. Then $\rho$ defines a linear form
$\tilde{\rho}\in\cD_c^*$ (where $\cD_c^*$ denotes the algebraic dual of
$\cD_c$) by means of\mfootnote{To define $\esp{A}_\rho$ with $\rho\in\cR$ and
  $A\in\cD_c$, $A$ is replaced by a Schwartz function differing from $A$
  outside of the support of $\rho$.}
\begin{equation}
\esp{\tilde{\rho},A} \equiv \esp{A}_\rho
.
\end{equation}
(We have used the notation $\esp{T,f}$ to denote the action of a linear form
$T$ on a vector $f$.) It should be clear that the linear map
$\hat\pi:\rho \to \tilde{\rho}$ from $\cR\to\cD_c^*$ is not one-to-one, as
there are many different $\rho$ yielding precisely the same expectation values,
and so the same linear map $\tilde{\rho}$.

Next, we can define the space $\cD_r$ as the span of Fourier modes $e^{ikx}$
on $\mS^1$. Clearly the analytic continuation operator $\cA$ is an isomorphism
of vector spaces from $\cD_r$ to $\cD_c$, namely, $A_c=\cA A_r$, with
$A_r(x)=\sum_k a_k e^{ikx}$ and $A_c(z)= \sum_k a_k e^{ikz}$. Therefore, the
dual spaces $\cD_r^*$ and $\cD_c^*$ are equally isomorphic. $P\in\cD_r^*$ can
then be defined as the linear form on $\cD_r$ matching $\tilde{\rho}$:
\begin{equation}
\esp{\tilde{\rho},\cA A_r} = 
\esp{P,A_r}
\qquad
\forall A_r\in \cD_r 
,
\end{equation}
that is, $P=\cA^T\tilde{\rho}$. The operator $K$, such that $P=K\rho$, is then
well-defined, and can be expressed as $K=\cA^T\hat{\pi}$.

It is noteworthy that even though $\rho$ is a distribution on the cylinder
$\mS^1\times\R$, the linear form $P$ needs not be a distribution (i.e., a
continuous linear form) on the circle $\S^1$. For instance, $\rho(z)
=\delta(z-z_0)$ (a two-dimensional Dirac delta) has expectation values
$\esp{e^{ikz}}_\rho = e^{ikz_0}$ and these are the Fourier components of
$P(x)$. When $\Im z_0\neq 0$ they are not polynomially bounded, hence $P$ is
not a Schwartz distribution on $\mS^1$. A simple way to choose the space $\cR$
so that the $P$ are bounded linear forms is to keep only the $\rho$'s which
contain a finite number of Fourier modes (with respect to $x\in\S^1$), each
mode weighted with a Schwartz distributions of bounded support with respect to
the variable $y$, i.e., $\rho=\sum_k\rho_k(y)e^{ikx}$ (a finite sum and
$\rho_k(y)$ of bounded support).

We have spelled out the definition of the operator $K$ in the setting of
periodic one-dimensional functions. Clearly the analogous constructions can be
carried out for more general compact groups using a decomposition in terms of
irreducible representations.

For sufficiently well behaved distributions $\rho$ on $\C^n$ the action of $K$
can be simply expressed as \cite{Okano:1992hp,Salcedo:1993tj}
\begin{equation}
P(\vx) = \int d^ny \, e^{ - i \, \vy \cdot \vnabla_x} \rho(\vx,\vy) 
\equiv \int d^ny \, \rho(\vx-i\vy,\vy) 
.
\label{eq:3.30}
\end{equation}
This is a straightforward consequence of
$\esp{A(x+iy)}_\rho = \esp{A(x)}_P$ for all $A$. \Eq{3.11} illustrates
this relation when $\rho(\vz)$ has the two-branch form in \Eq{3.10}.

\subsubsection{Construction of real representations from linearity}
\label{sec:III.D.2}

Let us assume that a complex density $P$ can be expressed as a linear
combination of some other densities $P_i$ 
\begin{equation}
P = \sum_i a_i P_i
,
\end{equation}
where the $a_i$ are some complex coefficients, with $\sum_ia_i=1$ if 
$P$ and the $P_i$ should be normalized. To avoid any convergence
issues we assume the collection $\{P_i\}$ to be finite. If each $P_i$ admits a
real representation $\rho_i$, $P_i= K \rho_i$, due to linearity of $K$, the
distribution
\begin{equation}
\ds \rho_c \equiv \sum_i a_i \rho_i
\end{equation}
will be a representation of $P$, i.e., $P=K\rho_c$.  Unfortunately, even if all
the $\rho_i$ a real, such $\rho_c$ will be complex in general since the $a_i$
are complex.

Abstracting what has been implicitly done in the previous subsections, in
order to obtain a real representation one can proceed as follows. 

First, the constant mode is treated separately and added a posteriori. So we
consider here complex distributions with {\em zero normalization}: $P = \sum_i
a_i P_i$ where the $P_i$, and hence $P$, integrate to zero.

Next, $K$ is a linear operator. Let us introduce the anti-analytic version of
$K$, which will be denoted by $\bar{K}$ and is also linear, through the
relation
\begin{equation}
\bar{K} \rho = (K\rho^*)^*
.
\end{equation}
Now given a collection of complex densities $P_i(x)$ we associate a set of
{\em complex} representations $\hat\rho_i(z)$ subject to the two (linear)
requirements
\begin{equation}
P_i = K \hat\rho_i
,\qquad
0 = \bar{K} \hat\rho_i
.
\label{eq:3.34}
\end{equation}
That is, the analytic projections of the $\hat\rho_i$ yield $P_i$ (i.e., the
$\hat\rho_i$ are representations of $P_i$ albeit complex) while their
anti-analytic projections vanish.
Then, obviously
\begin{equation}
\hat{\rho} \equiv \sum_i a_i \hat\rho_i
\end{equation}
is also a (complex) representation of $P$, i.e., $P=K\hat\rho$.

The second equation in \nec{3.34} is equivalent to
\begin{equation}
0 = K \hat\rho_i^*
.
\label{eq:3.34b}
\end{equation}
Hence
$0=K\hat\rho^*$, and
\begin{equation}
\rho  \equiv \hrho + \hrho^* =  2\Re(\hrho) 
\label{eq:3.37}
\end{equation}
is, by construction, a {\em real} representation of $P$,
\begin{equation}
 P = K \rho
.
\end{equation}

To finish the construction, the constant mode should be added to have properly
normalized distributions. Because the normalization of $P$ is real, its
constant mode, $P_0=1$, is real and it can be represented by a real $\rho_0$
which is added to $\hrho+\hrho^*$.

The two-branch construction follows the scheme of \Eqs{3.34} and these
equations admit many more solutions for a given collection $\{P_i\}$.  It is
interesting that unlike $\rho$, the complex representations $\hat\rho$ or
$\hat\rho_i$ preserve information on the phases of $P$ and $P_i$,
respectively. This implies that one can make new linear recombinations as long
as the complex representations are retained. This is no longer possible after
the real part operation is applied to obtain a real
representation.\mfootnote{And this is intriguingly similar to the problem of
  measurement and wave-function collapse in Quantum Mechanics.}

Another remark is that if $P_i$ has some symmetry, one can
impose the same symmetry on its complex representation $\hat\rho_i$, so each
symmetry type (irreducible representation of the symmetry group) can be
represented independently, thanks to the linearity of the construction.

The adaptation of this construction to the noncompact case deserves a separate
discussion. The expression in \Eq{3.6} holds equally well for a normalized
complex probability $P(x)$ defined on $\R$, using the Fourier components
$\tilde{P}(k)$ there,
\begin{equation}
P(x) = \int \frac{dk}{2\pi} \, e^{ikx}  \tilde{P}(k)
,\qquad
\int dx\, P(x) = 1
.
\end{equation}
The $k\to 0$ limits of $\tilde{Q}_\pm(k)$ in \Eq{3.6} exist, since
$\tilde{P}(0)$ is a real number. As a consequence $\tilde{Q}_\pm(0)$ take well
defined values, rather than being free parameters as in the compact case.

The functions $Q_\pm(x)$ receive (linear) contributions from $P(x)$ and
$P^*(x)$, and we can denote $\hat{Q}_\pm(x)$ the component coming only from $P$
(analogous to $\hat{\rho}$, as compared to $\rho=\hrho+\hrho^*$). In this case
one finds that the Fourier modes
\begin{equation}
\hat{\tilde Q}_\pm(k)
=
\pm \frac{ e^{\pm k Y}  }{2 \sinh(2kY)}  \tilde{P}(k)
,
\label{eq:3.40}
\end{equation}
display a pole at $k=0$.  This implies that the complex representations
$\hat{Q}_\pm(x)$ are not convergent at infinity. More precisely, their real
parts, $Q_\pm(x)$, are convergent but their imaginary parts are not.

In general, in the noncompact case, complex representations $\hrho_i$
corresponding to {\em normalized} $P_i$, will produce complex combinations
$\sum_i a_i \hat{\rho}_i$ which will not be properly convergent, however, the
divergence cancels in their real parts provided the normalization $\sum_i a_i$
is a real number.

Let us note that the infrared divergence must necessarily be present in
$\hat{\rho}$ (this is clear in \Eq{3.40}, since $\tilde{P}(0)=1$). This comes
from a conflict in \Eq{3.34} in the noncompact case. In the compact case, the
constant mode was cleanly separated and all distributions in \Eq{3.34} were
assumed to have zero normalization. The same cannot be done in the noncompact
case. If the constant mode cannot be extracted one finds an incompatibility in
\Eq{3.34}. To see this let us denote by $\hat{P}_0$ and $\hat{P}_{00}$ the
operators yielding the normalization of distributions on $\cM$ and $\cM^c$,
respectively ($\hat{P}_0=\int dx$ and $\hat{P}_{00}=\int d^2z$ for
$\cM=\R$). These operators fulfill the identities
\begin{equation}
\hat{P}_0 K = \hat{P}_0 \bar{K} = \hat{P}_{00}
.
\end{equation} 
Applying them to 
\begin{equation}
P = K \hrho
,\qquad
0 = \bar{K} \hrho
.
\label{eq:3.34a}
\end{equation}
one finds
\begin{equation}
1 = \hat{P}_0 P = \hat{P}_{00} \hrho
,\qquad
0 =  \hat{P}_{00} \hrho
.
\end{equation}
The conflict results in a singularity in the imaginary part of $\hrho$ at
the constant mode.

\section{\textsf{Localized representations on Lie groups}}
\label{sec:4}

In this section we aim at extending the previous constructions to non
necessarily abelian Lie groups. Eventually we will limit our study to compact
groups because too general (group) representations of noncompact groups would
be intractable, even qualitatively. Nevertheless, it can be conjectured that
our results apply also to a complex probability $P$ defined on any Lie group
$G$, provided $P$ is spanned by a set of well behaved representations of $G$
(e.g., bounded representations). The case $G=(\R^n,+)$ and $P(\vx)$ admitting
a Fourier decomposition in terms of $e^{i\vk\cdot\vx}$, for $\vk\in\R^n$ (as
opposed to $\vk\in\C^n$) is such an example.

\subsection{\textsf{Representations on groups}}

For definiteness we will assume a connected matrix group,
\begin{equation}
G = \{\, g(\va) = e^{\,\va \cdot \vT}, ~ \va\in\R^n \,\}
,
\end{equation}
where the matrices $\,T_i ~ (i=1,\ldots,n)$ are the group generators and $a^i
~ (i=1,\ldots,n)$ are the normal coordinates of the element $g$.  New
admissible real coordinate systems are derived by means of {\em real analytic}
changes of variables.

The complexified group $G^c$ is obtained by taking complex values for the
coordinates,
\begin{equation}
G^c = \{\, g(\va) = e^{\,\va \cdot \vT}, ~ \va \in \C^n \,\}
.
\end{equation}
The analytically extended observables are defined on $G^c$ through analytic
extension with respect to their dependence on the coordinates. (The extension
does not depend on the concrete coordinates used as long as they belong to the
class of admissible ones.)

Given a positive measure $d\mu(g)$ on $G$, one can define complex
distributions $P(g)$ on $G$ and corresponding expectation values. The factor
between two different choices of measure can be reabsorbed in the
distribution, so without loss of generality, we will use the right-invariant
Haar measure of $G$. For compact $G$ we adopt the normalized measure
\begin{equation}
\int_G dg = 1
\quad \text{(compact~$G$)}
\,.
\end{equation}
Likewise, we take the right-invariant measure on $G^c$. The complexified group
is never compact, but will be unimodular if $G$ is.\mfootnote{Since the
  invariant measure on $G^c$ is $|\sigma(\va)|^2 d^na\,d^na^*$ when the invariant
  measure on $G$ is $\sigma(\va)\,d^na$.} The concept of representation works as
before, as dictated by \Eq{2.2}.

We will need to introduce the (complex) conjugate element  $\bar{g}$ of a given
$g\in G^c$. This is defined by
\begin{equation}
g = g(\va),
\qquad
\bar{g} = g(\va^*)
\qquad 
\va \in \C^n
.
\end{equation}
This conjugation is a group automorphism in $G^c$ and its definition does not
depend on the particular coordinates used in $G$. Also note that $\bar{g}$
needs not coincide with $g^*$ (the conjugate matrix in a matrix group) unless
$\vT^*=\vT$.

An important property of the conjugation is that, for any (group)
representation $D(g)$ of $G$ and $D^*(g)=D(g)^*$ its conjugate representation,
upon analytic extension into $G^c$,
\begin{equation}
(D(g))^* = D^*(\bar{g})
\qquad
g \in G^c
.
\label{eq:4.11}
\end{equation}

Obviously, the set of autoconjugated (real) elements is $G$ itself,
\begin{equation}
g = \bar{g} 
\quad \text{iff} \quad g \in G
.
\end{equation}
The subset of {\em purely imaginary} elements of $G^c$, which we denote $G_I$,
can be naturally defined as
\begin{equation}
g \in G_I
\quad \text{iff} \quad 
 \bar{g} = g^{-1}
.
\end{equation}
In normal coordinates $G_I$ are those elements of $G^c$ with purely imaginary
coordinates. In the non abelian case $G_I$ is not a subgroup of $G^c$,
however if $g\in G$, $\, g G_I g^{-1} = G_I$. Also, if $h \in G_I$, $h^s\in
G_I$, for $s\in\R$.\mfootnote{For $G=\SU(2)$, the rotation group,
  $G^c=\SL(2,\C)$ is the Lorentz group and $G_I$ is the set of boosts.}
Furthermore, $G^c=GG_I=G_IG$.

\subsection{\textsf{Two-branch representations}}
\label{sec:4.2}

We will not need very general distributions on $G^c$, rather we use a
two-branch approach (with suitable variations in the higher dimensional case,
as in Sec. \ref{sec:3.2.2}). That is, for a given (normalized) complex
probability $P(g)$
\begin{equation}
1 = \int_G dg \, P(g) 
,
\end{equation}
we seek two positive distributions $Q_\pm(g)$ on $G$ in such a way that they
define a representation of $P(g)$, by means of the relation, analogous to
\nec{3.11},
\begin{equation}
P(g) = Q_+(g g_+) + Q_-(g g_-)
\qquad \forall g \in G
,
\label{eq:4.5}
\end{equation}
where $g_\pm \in G^c$ are two parameters of the construction, and
$Q_\pm(g g_\pm)$ refer to the analytic extension of $Q_\pm(g)$ into the
complexified group. Indeed, using the right-invariance of the measure,
\begin{equation}\begin{split}
\esp{A}_P 
&= \int_G dg \, A(g)  P(g) 
= \int_G dg \, A(g) \sum_{\sigma=\pm} Q_\sigma(g g_\sigma) 
\\
&= \int_G dg \sum_{\sigma=\pm} Q_\sigma(g) \, A(g g_\sigma^{-1}) 
\\ &
=
 N_+ \esp{A(g g_+^{-1})}_{Q_+} +  N_- \esp{A(g g_-^{-1})}_{Q_-}
,
\label{eq:4.6}
\end{split}\end{equation}
where $N_\pm$ denote the normalizations of $Q_\pm$,
\begin{equation}
N_\pm = \int_G dg \, Q_\pm(g)
,
\end{equation}
with
\begin{equation}
1 = N_+ + N_-,
\quad
N_\pm \ge 0
.
\end{equation}
\Eq{4.6} implies that the expectation value of $A$ can be obtained by
importance sampling of the two positive distributions $Q_\pm(g)$ defined on
$G$. The representation $\rho(g)$ itself has support on two copies of $G$
contained in $G^c$, namely, $Gg_+^{-1}$ and $Gg_-^{-1}$. Therefore the
elements $g_\pm$ represent the displacements away from $G$ into $G^c$.

In \Eq{4.5} we have arbitrarily chosen the shift to act on the right. Of
course everything would be analogous with $Q_+(g_+ g) + Q_-(g_- g)$. Also
possible would be (for a unimodular group)
\begin{equation}
P(g) = Q_+(g^\prime_+ g g_+ ) + Q_-(g_-^\prime g g_-)
.
\end{equation}
We do not explore this latter possibility as it is technically more
complicated with no obvious advantage.

It is clear that there is no solution to \Eq{4.5} (with positive $Q_\pm$) if
$g_\pm \in G$, unless $P$ is already a positive distribution. As discussed
in Sec. \ref{sec:2.c}, the representation $\rho(g)$ must have some support sufficiently far
from the real manifold (the group $G$ in this case); a minimal width is
required for any positive representation $\rho$.

The complex distribution $P$ is equivalent (has the same information as) to
two real functions, so it can be expected that for given $g_\pm$, the two real
functions $Q_\pm$ are essentially unique. To actually determine the two
branches $Q_\pm$ we apply the approach developed in Sec. \ref{sec:III.D.2} as
follows.

The (group) representations of a group span the space of complex functions
defined on that group (i.e., its regular representation
\cite{Barut:1986dd}). So general distributions $P(g)$ can be expanded as
linear combinations of (group) representations $D^R(g)$ of $G$, i.e., $P(g)
\sim \sum_R P^R D^R(g)$.

In order to cleanly separate the normalization mode (constant mode) in $P$, we
will assume in what follows that $G$ is a {\em compact group}, hence our
complex normalized probability $P(g)$ can be expressed as
\begin{equation}\begin{split}
P(g) &= 
1 + \sum_{R\not=1} \sum_{\alpha,\beta} (P^R){}^\beta{}_\alpha \, D^R(g)^\alpha{}_\beta
\\&
=
1 + \sum_{R\not=1} \tr( P^R  \, D^R(g))
.
\end{split}\end{equation}
The $P^R$ are constant complex matrices of the same dimension as the
representation $R$. We have separated the trivial (or singlet) representation
$D^{R=1}(g)\equiv 1$ which must carry weight $1$ if $P$ is normalized. 

As follows from the Peter-Weyl theorem, the set of irreducible representations
(irreps) form an orthonormal basis for the regular representation and we could
take the $R$ to be irreducible, however, such assumption is not strictly needed
for our construction, so we will only assume that $R$ does not contain the
trivial representations in its decomposition into irreps, therefore
\begin{equation}
\int_G dg \, D^R(g) = 0
\qquad (R\not=1)
.
\end{equation}

To apply the scheme of Sec. \ref{sec:III.D.2}, we will seek {\em complex}
representations for each component $R$ in $P$, fulfilling the conditions in
\Eqs{3.34}. That is, for each $R\not=1$ we seek two functions
$\hat{Q}^R_\pm(g)$ of the form
\begin{equation}
\hat{Q}^R_\pm(g) = \tr( Q^R_\pm \, D^R(g))
,
\label{eq:4.16}
\end{equation}
where $Q^R_\pm$ are two matrices to be determined. Then the real distributions
\begin{equation}
Q^R_\pm(g) = 2\Re \, \hat{Q}^R_\pm(g) 
\qquad
g \in G
\end{equation}
are the two real branches in the representation of the component $R$ of
$P(g)$ and
\begin{equation}
Q_\pm(g) = N_\pm + \sum_{R\not=1} Q^R_\pm(g)
.
\end{equation}

The two functions $\hat{Q}^R_\pm(g)$ are to be determined through
\Eq{3.34}. The action of the operator $K$ in our case can be read off from
\Eq{4.5} since that equation is just $P=K\rho$.

The representation condition on $\hat{Q}^R_\pm(g)$ (first relation in
\Eq{3.34}) becomes (using \Eq{4.16})
\begin{equation}\begin{split}
\tr(P^R D^R(g))
&=
\sum_{\sigma=\pm}\hat{Q}^R_\sigma (g g_\sigma) 
=
\sum_{\sigma=\pm}
\tr(Q^R_\sigma \, D^R(g)D^R(g_\sigma)) 
,
\end{split}\end{equation}
that is
\begin{equation}
P^R = D^R(g_+) \, Q^R_+ + D^R(g_-) \, Q^R_-
.
\label{eq:4.19}
\end{equation}

To impose the second relation in \Eq{3.34}, note that
\begin{equation}
\hat{Q}^R_\pm(g)^* = \tr( Q^R_\pm{}^* \, D^{R*}(g))
,\quad g\in G
,
\end{equation}
where $D^{R*}$ is the conjugate representation of $D^R$. Then
\Eq{3.34b} takes the form
\begin{equation}
0 = \sum_{\sigma=\pm}
\tr\left( Q^R_\sigma{}^* \, D^{R*}(g) \, D^{R*}(g_\sigma) \right) 
.
\end{equation}
Taking complex conjugation and using \Eq{4.11} yields
\begin{equation}
0 = \sum_{\sigma=\pm}
\tr\left( Q^R_\sigma \, D^R(g)D^R(\bar{g}_\sigma) \right)
,
\end{equation}
which provides a second equation on $Q^R_\pm$:
\begin{equation}
0 = 
D^R(\bar{g}_+) \, Q^R_+ + D^R(\bar{g}_-) \, Q^R_-
.
\label{eq:4.23}
\end{equation}

Assuming that the required matrices are invertible, the system of \Eqs{4.19}
and (\ref{eq:4.23}) can be solved to give
\begin{equation}
Q^R_\pm = \left( 
D^R( g_\mp^{-1} g_\pm ) - D^R( \bar{g}_\mp^{-1} \bar{g}_\pm )
\right)^{-1} D^R ( g_\mp^{-1} ) \, P^R .
\label{eq:4.24}
\end{equation}

Equivalently,
\begin{equation}
Q^R_\pm = \left( 1 -
 D^R( g_\pm^{-1} g_\mp \bar{g}_\mp^{-1} \bar{g}_\pm )
\right)^{-1} D^R ( g_\pm^{-1} ) \, P^R .
\label{eq:4.25}
\end{equation}
So a solution is obtained whenever the matrix $D^R( g_\pm^{-1} g_\mp
\bar{g}_\mp^{-1} \bar{g}_\pm ) $ has no eigenvalue $\lambda=1$. If it has,
there can still be solutions if $P$ has no component along those
eigenvectors. We come back to this crucial question in Sec. \ref{sec:5}.
For the time being we will assume that the required matrices are indeed
invertible. As always the trivial representation (constant mode) has been
explicitly extracted (since certainly all eigenvalues $\lambda=1$ when $R=1$).

As noted $G^c=GG_I$. Since the factors of $g_\pm$ along $G$ are
ineffective, the most efficient choice, in principle, corresponds to taking
purely imaginary displacements. Hereafter we adopt this prescription, $g_\pm
\in G_I$, and also choose a symmetric disposition of the two shifts, $g_+ =
g_-^{-1}$:
\begin{equation}
h \equiv g_+ = g_-^{-1} = \bar{g}_+^{-1} = \bar{g}_- \in G_I
.
\label{eq:4.26}
\end{equation}
Then \Eq{4.5} becomes
\begin{equation}
P(g) = Q_+(g h) + Q_-(g h^{-1})
\qquad \forall g \in G
,
\label{eq:4.5a}
\end{equation}
and
\begin{equation}
\esp{A}_P =
 N_+ \esp{A(g h^{-1})}_{Q_+} +  N_- \esp{A(g h)}_{Q_-}
.
\label{eq:4.6a}
\end{equation}
Also \Eqs{4.19} and (\ref{eq:4.23}) become
\begin{equation}\begin{split}
P^R &= D^R(h) \, Q^R_+ + D^R(h^{-1}) \, Q^R_-
\\
0 &= D^R(h^{-1}) \, Q^R_+ + D^R(h) \, Q^R_-
\end{split}
\,.
\label{eq:4.23b}
\end{equation}
In addition \Eq{4.24} becomes
\begin{equation}
Q^R_+ = \chi(D^R(h)) \, P^R
,\quad
Q^R_- = \chi(D^R(h^{-1})) \, P^R
,
\label{eq:4.31a}
\end{equation}
where $\chi$ is the function introduced in \Eq{3.18} and $\chi(D^R(h))$ is a
matrix of the same dimension as $R$. Therefore, the two branches for the
representation of $P(g)$ can be compactly written as
\begin{equation}
Q_\pm (g) = N_\pm  + 2\Re \sum_{R\not=1} \tr \!\left[
P^R \,
D^R(g) \, \chi( D^R(h^{\pm 1}) )
\right]
.
\label{eq:4.30}
\end{equation}

Because $G$ is compact and its representations $R$ are unitary, the matrices
$D^R(g)$ are unitary, while $D^R(h)$ (and hence $\chi(D^R(h))$) are
hermitian. This follows from the identity
\begin{equation}
D^R(g^{-1})  = D^R(g)^{-1}  = D^R(\bar{g})^\dagger
\quad \forall g \in G^c
\quad\text{($R$ unitary)}
.
\label{eq:4.31}
\end{equation}

Once again, for sufficiently large $h$ (assuming no $\lambda=1$ eigenvalues
are involved) $\chi(D^R(h) )$ goes to zero and only the singlet (trivial
representation) mode remains in \Eq{4.30}, implying that eventually $Q_\pm$
become non negative.

Of course the case $G=\U(1)^{\times n}$ studied in Sec. \ref{sec:3} conforms
to this general scheme: the normal coordinates are $\va=\vx$ in $G$ and
$\va=\vz$ in $G^c = (\U(1)\times\R)^{\times n}$. Also, $R=\vk$, $D^R(g) =
e^{i\vk\cdot\vx}$ and $P^R= \tilde{P}_\vk$. Furthermore, $h$ has coordinates
$-i\vY$ and so $D^R(h) = e^{\,\vk\cdot\vY}$. In this way \Eq{4.30} reproduces
\Eq{3.17}.

\subsection{\textsf{An $\SU(2)$ example}}
\label{sec:4.3}

Let us consider the following complex probability on
$G=\SU(2)$
\begin{equation}
P(g) = 1 + \tr(pg)
\,
\quad
g\in \SU(2)
.
\label{eq:4.37}
\end{equation}
Here $p$ is a constant complex $2\times 2 $ matrix. Letting $h\in \SU(2)_I$, a
direct application of the previous results gives
\begin{equation}
Q_\pm(g) = N_\pm 
\pm 2\Re \,\tr \left( h^{\pm 1}(h^2-h^{-2})^{-1} p g \right)
.
\end{equation}

To be more explicit, let
\begin{equation}\begin{split}
g &= \cos(\psi/2) - i\sin(\psi/2) \, \hat\vpsi\cdot \vsigma
,
\\
p &= p_0  + \vp \cdot \vsigma
,
\\
h &= \cosh(Y) + \sinh(Y) \, \hat\vY\cdot \vsigma,
\end{split}\end{equation}
where $p_0$ and $\vp$ can be complex and $\vpsi$ and $\vY$ are real. Then
\begin{equation}
Q_\pm(g) = A_\pm a_0 + \vB_\pm \cdot \va 
\end{equation}
with
\begin{equation}
a_0 = \cos(\psi/2) , \quad
\va = \sin(\psi/2)\hat\vpsi
\end{equation}
and
\begin{equation}\begin{split}
A_\pm &=
\frac{\Re (p_0)}{\cosh (Y)} \pm \frac{\hat\vY\cdot\Re(\vp)}{\sinh (Y)} 
\\
\vB_\pm &=
\pm \frac{\hat\vY \, \Im (p_0)}{\sinh (Y)} 
\pm \frac{\hat\vY \times \Re(\vp) }{\sinh (Y)} 
+ \frac{\Im(\vp)}{\cosh (Y)} 
.
\end{split}\end{equation}

\begin{figure}[t]
\begin{center}
\includegraphics[height=50mm]{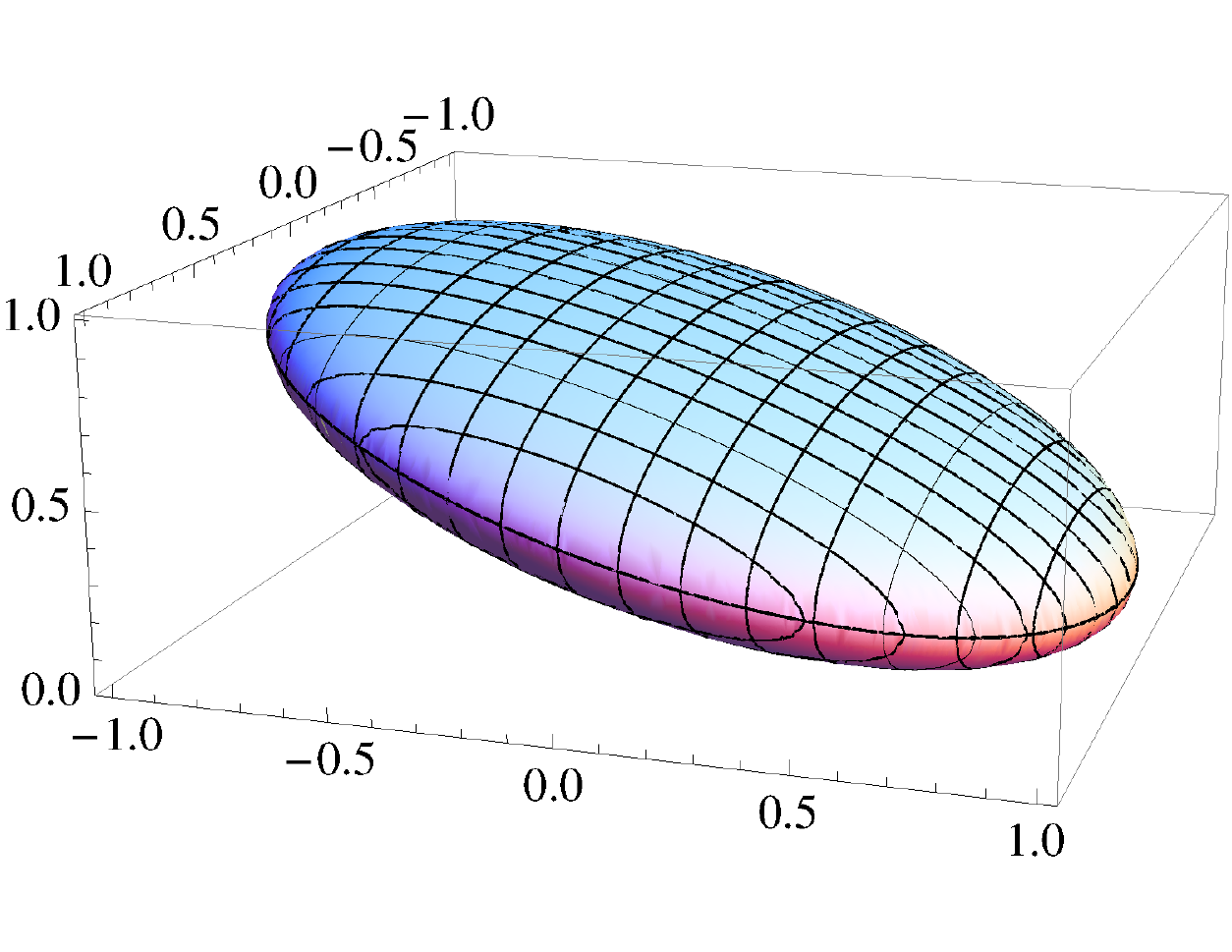}
\end{center}
\caption{For $g\in\SU(2)$, function $Q_+(g)$ on the plane $(a_1,a_3)$ with
  $a_2=0$, for $P(g) = 1+\beta \, \tr(g)$ with $\beta=1-i$, and
  $h=\diag(e^Y,e^{-Y})$ with $Y=3.5$, and $N_+=1/2$. }
\label{fig:2}
\end{figure}
As an illustration, in Fig. \ref{fig:2} we show the function $Q_+(g)$ for
\begin{equation}
P(g) = 1 + \beta \, \tr(g)
,
\qquad
\beta = 1-i
,
\end{equation}
using $\hat\vY = (0,0,1)$ and $Y=3.5$, and $N_+=\h$. $\SU(2)$ is a
three-sphere, $a_0^2+\va^2=1$, so $Q_+$ as a function of $\va$ is two-valued.
The plot displays $Q_+(a_1,0,a_3)$, the submanifold $a_2=0$ being a
two-sphere.

It is interesting to note that in any $\U(N)$ group the complex probabilities
of the type in \Eq{4.37} 
can be reduced to a standard form before representation. The matrix $p$ can be
written as
\begin{equation}
p  = u_L d u_R,
\quad
u_{L,R} \in \U(N),
\quad d~\text{diagonal and non negative}
,
\end{equation}
so that
\begin{equation}
P(g) = 1+ \tr(d u_R g u_L)
.
\end{equation}
Then it is sufficient to find representations $\rho^\prime(g)$ for
\begin{equation}
P^\prime(g)  =  1 +\tr(d g)
,
\end{equation}
and afterwards undo the left and right translations
\begin{equation}
\rho(g) = \rho^\prime( u_R g u_L)
.
\end{equation}
In the case of $\SU(N)$,
\begin{equation}
p  = e^{i\varphi} u_L d u_R,
\quad
u_{L,R} \in \SU(N),
\quad d~\text{diagonal and non negative}
.
\end{equation}
($\varphi$ real.)
In particular for $\SU(2)$ the most general case required is
$p = e^{i\varphi} a(1+\cos\theta \sigma_3)$, $a>0$, ~$\theta,\varphi \in\R$.

\subsection{\textsf{Representations through convolutions}}

The functions $Q_\pm(g)$ can also be obtained from convolution of $P(g)$ with a
fixed kernel. To do this, we express $P$ in terms of irreducible group
representations, $\mu$, as
\begin{equation}\begin{split}
P(g) &= \sum_\mu \tr( P^\mu D^\mu(g))
,
\\
 P^\mu &= n_\mu \int_G dg \, P(g) \, D^\mu(g^{-1})
,
\end{split}\end{equation}
where $n_\mu$ denotes the dimension of the irrep $\mu$.  Using the expression
of $P^\mu$ to work out \Eq{4.30}, one obtains\mfootnote{The group
  convolution \[(A*B)(g) \equiv \int_G dg^\prime\,
  A(g^\prime)B(g^\prime{}^{-1} g) = \int_G dg^\prime\, A(g
  g^\prime{}^{-1})B(g^\prime) \] is not commutative in general.}
\begin{equation}\begin{split}
Q_\pm (g) &= N_\pm + 2\Re \int dg^\prime \,  P(g^\prime) 
\, C(g^\prime{}^{-1} g \,; \, h^{\pm 1})
\\ & =
N_\pm + 2\Re \big(  P(g) * C(g \,; \, h^{\pm 1}) \big)
,
\label{eq:4.29}
\end{split}\end{equation}
with
\begin{equation}
C(g; h) = \sum_{\mu\not=1} n_\mu \,\tr \big[ D^\mu(g) \, \chi(D^\mu(h) ) \big]
.
\label{eq:4.33}
\end{equation}
\Eqs{4.29} and (\ref{eq:4.33}) generalize \Eqs{3.20} and (\ref{eq:3.19}),
respectively.

Summations on $\mu$ or $\mu^*$ (the conjugate irrep) are equivalent within the
trace in \Eq{4.33}. Using this rearrangement, along with \Eq{4.31} and
\begin{equation}
D^{R*}(g) = D^R(g^{-1})^T
\quad \forall g \in G^c
\quad\text{($R$ unitary)}
\end{equation}
one can easily establish the following identities
\begin{equation}
C(g; h^{-1}) = C(g^{-1} ; h) =
C(g;h)^*
.
\end{equation}

\subsection{\textsf{Representations in matrix groups}
\label{sec:4.5}}

Let $G \subseteq \U(N)$, and $g^i{}_{\!j}$ the matrix elements of $g\in G$
~($i,j=1,\ldots N$). The (group) representations of $G$ can be obtained from
tensor product of the basic representations $g$ and $g^*$. (Note that such
product representations will be reducible in general.)

In the simplest case in which only $g$ is involved
\begin{equation}
P(g) = 
\sum_{n=0}^\infty
p^{j_1\ldots j_n}_{\, i_1\ldots \, i_n}
\, g^{i_1}{}_{\!j_1} \cdots g^{i_n}{}_{\!j_n} 
,
\label{eq:4.49}
\end{equation}
where $p^{j_1\ldots j_n}_{\, i_1\ldots \, i_n}$ are complex coefficients. This
is a decomposition of $P$ into group representations of the type
$D^R(g) = g\otimes \cdots \otimes g $ ($n$ factors),
\begin{equation}
D^R(g)_{j_1\ldots j_n}^{\, i_1\ldots \, i_n}
=
g^{i_1}{}_{\!j_1} \cdots g^{i_n}{}_{\!j_n}
\equiv
(g^{\otimes n} )_{j_1 \ldots j_n}^{\, i_1\ldots \, i_n}
,
\end{equation}
and \Eq{4.30} applies
\begin{equation}
Q_+(g) = N_+ + 2\Re \!\sum_{n=1}^\infty \hat{Q}_n(g)
,
\label{eq:4.53}
\end{equation}
with
\begin{equation}
\hat{Q}_n(g) = 
p^{j_1\ldots j_n}_{\, i_1\ldots \, i_n}
\, (g^{\otimes n} )_{k_1 \ldots \, k_n}^{\, i_1\ldots \, i_n}
\, \chi(h^{\otimes n} )^{k_1 \ldots \,k_n}_{j_1\ldots j_n}
.
\end{equation}
The contribution to $Q_-(g)$ is analogous, using $h^{-1}$ instead of $h$.
Also note that because $R$ is unitary, $h$ is hermitian.

Let us assume that $h \in G_I$ is a diagonal matrix,
\begin{equation}
h = \diag( \omega_1,\ldots,\omega_N )
.
\end{equation}
The $\omega_{\,i}$ are real (and moreover positive for a connected group). In
this case $h^{\otimes n}$ and $\chi(h^{\otimes n})$ are also diagonal and
$\hat{Q}_n(g)$ takes a simple form
\begin{equation}
\hat{Q}_n(g) = 
p^{j_1\ldots j_n}_{\, i_1\ldots \, i_n}
\, g^{i_1}{}_{j_1} \cdots g^{i_n}{}_{j_n} 
\, \chi(\Omega)
,
\quad
\Omega = \omega_{j_1}\cdots \omega_{j_n}
.
\label{eq:4.54}
\end{equation}
$\Omega$ denotes the argument of the function $\chi$ generated by the
displacement to the complex manifold. We can see that $\Omega$ picks up a
factor $\omega_j$ for each factor $g^i{}_j$ in the representation $R$.

More generally, $R= g^{\otimes n} \otimes g^*{}^{\otimes m}$. The
corresponding right translation with $h\in G_I$ is
\begin{equation}
g \mapsto g h 
\qquad
g^* = g^{-1T} \mapsto  g^* h^{-1T}
.
\end{equation}
This implies that $\Omega$ picks up a factor $\omega_j$ for each factor
$g^i{}_j$ in $R$, and a factor $\omega^{-1}_j$ from each factor
$g^*{}^i{}_j$. That is, a term
\begin{equation}
P(g) = 
\, g^{i_1}{}_{\!j_1} \cdots g^{i_n}{}_{\!j_n} 
\, g^*{}^{l_1}{}_{\!k_1} \cdots g^*{}^{l_m}{}_{\!k_m} 
,
\end{equation}
gives a contribution
\begin{equation}\begin{split}
\hat{Q}(g) &= 
\, g^{i_1}{}_{\!j_1} \cdots g^{i_n}{}_{\!j_n} 
\, g^*{}^{l_1}{}_{\!k_1} \cdots g^*{}^{l_m}{}_{\!k_m} 
\, \chi(\Omega)
,
\\ 
\Omega &= \omega_{j_1}\cdots \omega_{j_n}
\omega^{-1}_{k_1}\cdots \omega^{-1}_{k_m}
.
\label{eq:4.57}
\end{split}\end{equation}
Similar formulas hold in more general cases.\mfootnote{In $\GL(N,\C)$, a non
  compact group, $R$ would be obtained as a direct product of basic
  representations $g$, $g^*$, $g^{-1T}$ and $g^{-1\dagger}$. The
  corresponding right translation with $h\in G_I$ (which is no longer
  hermitian) would be $gh$, $g^*h^{-1*}$, $g^{-1T} h^{-1T}$ and $g^{-1\dagger}
  h^\dagger$, respectively.  So, for instance, a term of the form
$
P(g) = 
\, g^{i_1}{}_{\! j_1} \, g^*{}^{i_2}{}_{\! j_2}
 (g^{-1T})^{i_3}{}_{\! j_3}  (g^{-1\dagger}){}^{i_4}{}_{\! j_4}
$
would produce a contribution
$
\hat{Q}(g) = 
\, g^{i_1}{}_{\! j_1} \, g^*{}^{i_2}{}_{\! j_2}
 (g^{-1T})^{i_3}{}_{\! j_3}  (g^{-1\dagger}){}^{i_4}{}_{\! j_4}
\, \chi(
\omega_{j_1} \omega_{j_2}^{-1*} \omega_{j_3}^{-1} \omega_{j_4}^*
)
.
$
} Also note that $g^{\otimes n}$ suffices for $\SU(2)$ since $g$ and $g^*=
\sigma_y g \sigma_y $ are equivalent representations in this case.

Another observation is that $\Omega$ may be equal to $1$ for some components
and the previous formulas do not directly apply there. This will certainly
happen when $R$ contains the trivial representation in its reduction, but not
only then. This problem is addressed in Sec. \ref{sec:5}.

If a configuration of the real manifold consists of $n$ variables,
$(\overset{1}{g},\ldots, \overset{n}{g}) $, each of them an element of the
group $G_1 \subseteq \U(N)$, the complex probability is defined on the
group $G = G_1\times \cdots \times G_1$ ($n$ factors) and $g=\overset{1}{g}
\cdots \overset{n}{g} $. The formulas apply as before, and for instance, a
term of the form
\begin{equation}
P(g) = 
\overset{1}{g}{}^{\,i_1}{}_{\!j_1}
\, \overset{1}{g}{}^*{}^{\,i_2}{}_{\!j_2} 
\, \overset{2}{g}{}^{\,i_3}{}_{\!j_3} 
,
\label{eq:4.60}
\end{equation}
with diagonal $h$ with parameters $\overset{r}{\omega}{}_j$, would yield a
contribution
\begin{equation}
\hat{Q}(g) = 
\overset{1}{g}{}^{\,i_1}{}_{\!j_1}
\, \overset{1}{g}{}^*{}^{\,i_2}{}_{\!j_2} 
\, \overset{2}{g}{}^{\,i_3}{}_{\!j_3} 
\, \chi(
\overset{1}{\omega}{}_{j_1}
\, \overset{1}{\omega}{}^{-1}_{j_2}
\, \overset{2}{\omega}{}_{j_3}
)
.
\label{eq:4.61}
\end{equation}

It should be noted that a discussion similar to that in Sec. \ref{sec:3.2.2}
can be (and should be) done here to restore uniformity with respect to the $n$
variables, resulting in a total of $2^n$ branches, instead of $2$. An explicit
nonabelian example using $2^n$ branches is analyzed in Sec. \ref{sec:5.c}. In
the abelian case, bifurcation of the variables solved the problem of singular
terms ($\vk\cdot\vY=0$ denominators). A crucial difference with the abelian
case is that the presence of singular components (not invertible matrices in
\Eq{4.30}) is not automatically solved by bifurcation in the nonabelian case,
so we defer the discussion to Sec. \ref{sec:5}.

When the element $h \in G_I$ is not directly diagonal but it is diagonalizable
within $G$,\mfootnote{When $G \subset \U(N)$ the elements are diagonalizable,
  but not all elements need to have a diagonal representative in their
  conjugacy class. That is, their diagonal version may lie outside $G$. A
  similar consideration holds for $G_I$.} a practical way to proceed is as
follows. Let
\begin{equation}
h = U h_z U^{-1}
,\qquad
U\in G,
\quad
h_z \in G_I \text{~and diagonal}
,
\end{equation}
and let
\begin{equation}
P^\prime(g) \equiv P( U g \, U^{-1})
.
\end{equation}
Then
\begin{equation}
\hat{Q}(g) = 
\hat{Q}^\prime(U^{-1} g \, U)
,
\label{eq:4.62}
\end{equation}
where $\hat{Q}^\prime(g)$ is the complex representation associated to
$P^\prime(g)$, constructed using the diagonal $h_z$ as described
above. Indeed, using \Eq{4.30},
\begin{equation}\begin{split}
\hat{Q}(g) &= 
\tr\left(
P^R  D^R(g)\chi(D^R(h))
\right)
\\ &
=
\tr\left(
P^R  D^R(g) D^R(U)\chi(D^R(h_z))D^R(U)^{-1}
\right)
\\ &=
\tr\left(
P^\prime{}^R D^R(U^{-1} g \, U)\chi(D^R(h_z))
\right)
=
\hat{Q}^\prime(U^{-1} g \, U)
.
\end{split}\end{equation}

\section{\textsf{Removal of singular kernels and examples}}
\label{sec:5}

\subsection{\textsf{Singular kernels}}
\label{sec:5.a}

The first expression in \Eq{4.31a} can be rewritten as
\begin{equation}
Q^R_+ = D^R(h)^3
\left( D^R(h)^4  - 1 \right)^{-1} P^R 
\label{eq:4.65}
\end{equation}
and similarly for $Q^R_-$ with $h^{-1}$. Hence there is a proper solution when
$D^R(h)$ has no $\lambda=1$ eigenvalues\mfootnote{If $D^R(h)$ has no unit
  eigenvalue $D^R(h^4)$ could still have it but this can be circumvented by
  considering another element $h^s$ with suitable real $s$ (analogous to a
  change in the parameter $Y$ before). What really matters is the
  uniparametric subgroup $H=\{ h^s, \, s \in \R \}$, or equivalently the Lie
  algebra generator $t$ of $h=e^t$. Unit eigenvalues of $D^R(h)$ match to zero
  eigenvalues of $t$ in the representation $R$.} or, if it has, $P^R$ has no
components along the corresponding eigenvectors. Otherwise we meet an
obstruction to solving \Eq{4.23b}.

As already noted, when a probability $P(g)$ is complex, the support of any of
its {\em real} representations must necessarily extend beyond $G$ into the
complexified manifold. In the two-branch approach the pushing into $G^c$ is
carried out by $h$ (or more generally $g_\pm$). An obstruction arises when
some components of $P^R$ are not moved by $D^R(h)$ (unless they happen to be
already  positive). The obstruction takes
place when some components of $P$ remain invariant under the action of $h$ ,
i.e., when $h$ does not act {\em effectively} on all components of $P$. This
is quite clear in the abelian case $\U(1)^{\times  n}$ discussed in
Sec. \ref{sec:3.2.1}. There, an obstruction was met for Fourier modes such
that $\vk\cdot\vY=0$. They correspond to the the Fourier components
$e^{i\vk\cdot\vx}$ of $P$ which remain invariant under the imaginary
translation $\vx\to\vx-i\vY$.

An important observation is that, in the nonabelian case, the obstruction
cannot be removed by a clever choice of $h$ (or even $g_\pm$ outside
$G_I$). To see this it suffices to consider the case $G=\SU(2)$. If $R=j$ is a
half-integer representation, $D^j(h)$ has no eigenvalue equal to $1$, since
the operator $J_z$ has no zero eigenvalues, and the same is true of
$J_n=\hat{\vn}\cdot\vJ$; so for those irreps any choice of rotation axis
provides a solution.\mfootnote{In the $\SU(2)$ example discussed in
  Sec. \ref{sec:4.3}, besides the trivial representation, only $j=1/2$ was
  involved, so no obstruction arose in that case.} However, for integer $j$,
$J_n$ has exactly one zero eigenvalue. This means that no matter how the
(complex) rotations are chosen $D^j(h)$ will have an eigenvalue equal to one
for some eigenvector. We conclude that for integer $j$ the obstruction cannot
be avoided by just a better choice of the element $h$.  For $h$ imaginary the
rotation angle is imaginary and the rotation axis $\hat{n}$ is real. Choosing
a complex axis\mfootnote{For $g_\pm\in G^c$ \Eq{4.24} generalizes \Eq{4.31a}.}
would not help though: if $J_n$ has a zero eigenvalue whenever $\hat{n}$ is
real (and so $\det(\hat{n}\cdot J)=0$) by analytic extension, the zero will
persist in the complex case too. Thus we stick to the choice $h\in G_I$.

It follows that for certain groups and representations there is no perfect
choice of a single $h$ that would work simultaneously for all components of a
general complex probability $P$. The obvious solution is to try to decompose
$P$ as a sum of terms in such a way that each term can be treated effectively
by a different suitable element $h$:
\begin{equation}
P(g) = 1 + \sum_{k=1}^m P_k(g)
\qquad
\text{($h_k \in G_I$ and acts effectively on $P_k$)}
.
\label{eq:5.2}
\end{equation}
\Eq{4.65} would then apply for each term $k=1,\ldots,m$ separately without
obstruction, and each $h_k$ would introduce a further pair of branches in the
support of $\rho$. The arguments given at the end of Sec. \ref{sec:3.1}
indicate the number $m$ of terms should be as small as possible.

In a setting like that of \Eq{4.57}, i.e., a matrix group with diagonal $h$,
the obstruction appears for those components with
$\Omega=1$.\mfootnote{Throughout $\Omega$ denotes a generic argument of the
  function $\chi$, e.g. in \Eq{4.57}. $\Omega$ is any of the eigenvalues of
  $D^R(h^{\pm 1})$.} A simpleminded approach would be to use such diagonal $h$
for the $\Omega\not=1$ terms and a different element $h^\prime$ for the
remainder. However such strategy is not practical in general. To see this
consider again $\SU(2)$ and a representation $R=j$, with integer $j$ (since
the half-integer irreps pose no problem).  A diagonal $h=h_z$ corresponds to a
rotation around the $z$ axis. The components in $P$ can be decomposed in the
$J_z$ basis $|j,m\rangle_z$, and $|j,0\rangle_z$ will be unaffected by
$h_z$. A simple prescription is to identify such components from the condition
$\Omega=1$. All the $\Omega\not=1$ terms can be treated with $h_z$ (of
sufficient magnitude to guarantee positivity of the representation). The terms
with $\Omega=1$ should be treated with a different element $h_n$,
corresponding to a rotation around some axis $\hat\vn$. As it turns out, one
cannot take just any axis. The reason is that we need $h_n$ to act effectively
on $|j,0\rangle_z$: This vector can be decomposed in the basis $|j,m\rangle_n$
and one should take $\hat\vn$ in such a way that $|j,0\rangle_z$ has no
component along $|j,0\rangle_n$ (since such component would remain unaffected
by $h_n$). Hence, the axis $\hat\vn$ must fulfill the condition
\begin{equation}
{}_z\langle j,0 | j,0 \rangle_n = 0
.
\label{eq:5.3}
\end{equation}
In practice, this means that the cosine of the angle between the $z$ axis and
$\hat\vn$ should be a zero of the $j$-th Legendre polynomial,
$P_j(\hat{e}_z\cdot\hat{\vn})=0$.  For all odd $j$, $\vn=\hat{e}_x$ suffices.
Unfortunately for even $j$ the axis must be changed for different $j$ and in
general an infinite number of branches could be required.

So a method is needed to implement \Eq{5.2} using a common (and small) set of
branches for all representations simultaneously. This can be done as follows.

Let the set of elements $h_k\in G_I$, $k=1,\ldots,m$, where the number $m$ is
to be chosen appropriately for the given group. For any irrep $R\not=1$, let
$V^R$ be the $n_R$-dimensional vector space where $D^R(g)$ acts ($n_R=\dim
R$). Each $h_k$ defines a {\em singlet} subspace $W^R_k$ of $V^R$ (which may
be $\{0\}$); singlet means that within this subspace $h_k$ acts as the
identity operator:
\begin{equation}
W^R_k = \{ v\in V^R, \, D^R(h_k) v= v \}
.
\end{equation}
On the orthogonal complement $W^{R,\perp}_k$ the element $h_k$ acts
effectively [i.e., no non null vector of $W^{R,\perp}_k$ is left invariant by
$D^R(h_k)$] and
\begin{equation}
V^R = W^R_k \oplus W^{R,\perp}_k
.
\end{equation}
The obstruction is avoided for the irrep $R$ if any vector of $V^R$
can be decomposed as a sum where each term is acted
effectively by $h_k$, i.e.
\begin{equation}
\forall v\in V^R
\qquad
v=\sum_{k=1}^m v_k
,
\qquad
v_k\in W^{R,\perp}_k
\,.
\end{equation}
In other words,
\begin{equation}
V^R = W^{R,\perp}_1 + \cdots + W^{R,\perp}_m
.
\label{eq:4.69}
\end{equation}
(This is the plain sum of subspaces, no mutual null intersection nor
orthogonality is assumed.) If \Eq{4.69} holds for a {\em fixed set of $h_k$
  common to all irreps $R$}, the complex probability representation problem is
solved for the group.  Note that the $P^R$ appearing in the decomposition of
$P$ are matrices rather than vectors of $V^R$, however, since $h$ acts on the
left [e.g. \Eq{4.65}] one can view $P^R$ as a set of $n_R$ column vectors of
$V^R$ and apply the method to these vectors, then $P^R$ gets decomposed as a
sum of matrices each one acted effectively by one of the $h_k$, as required in
\Eq{5.2}. The decomposition $v=\sum_k v_k$ is not unique in general and so
some canonical prescription can be adopted to fix the ambiguity.

Now let us show that suitable sets of elements $\{ h_k\in G_I, \, k=1,\ldots,
m \}$ do exist for any Lie group $G$. Let us write $h_k=e^{it_k}$ where $t_k$
are in the Lie algebra of $G$. A {\em sufficient} condition to fulfill
\Eq{4.69} simultaneously for all irreps $R$ is that the $t_k$ generate the Lie
algebra, or equivalently, the elements $e^{t_k}$ generate $G$.\mfootnote{I.e.,
  the minimal algebra containing $\{t_k,\,k=1,\ldots,m\}$ is the whole
  algebra, and the minimal subgroup containing all the subgroups
  $\{e^{st_k},\,s\in\R\}$ is $G$ itself.} To see that this is sufficient, let
us first note that the condition \Eq{4.69} is equivalent to
\begin{equation}
W^R_1 \cap \cdots \cap W^R_m
= \{0\}
.
\label{eq:4.69a}
\end{equation}
This follows from the property $(A+B)^\perp = A^\perp \cap B^\perp$ and the
fact that the spaces are finite-dimensional (hence $A^{\perp\perp}=A$)
\cite{Halmos:1974}. The equivalence implies that (upon suitable decomposition)
the set of elements $h_k$ acts {\em effectively} on any vector of $V^R$
[\Eq{4.69}] if and only if there are no nontrivial singlet vectors common to
all the $h_k$ simultaneously [\Eq{4.69a}]. But the latter condition is
guaranteed if the $e^{t_k}$ generate $G$. Indeed, let us assume that there
were a non trivial singlet $|s\rangle$ common to all the $t_k$, i.e.,
$D^R(h_k)|s\rangle=|s\rangle$. Then the stability group of $|s\rangle$ would
contain all the $e^{t_k}$ and so it would coincide with $G$. This would imply
that $V^R$ contains a proper invariant subspace (namely the multiples of
$|s\rangle$) in contradiction with the assumption that $R \not = 1$ is
irreducible.

We have just shown that if the set $\{ t_k, \, k=1,\ldots,m \}$ generates the
whole Lie algebra, any $P^R$ can be decomposed as a sum of terms in such a way
that at least one of the $h_k$ acts effectively on each term, and this for all
the irreps $R$ except the trivial one. Certainly, if one takes as $t_k$ all
the elements of a linear basis of the algebra, they generate the whole
algebra, so it is never necessary to take $m$ larger than $n$ ($n$ being the
dimension of the group $G$) and in general a smaller $m$ is sufficient.

The condition that the set of elements $t_k$ must generate the whole algebra
is sufficient but certainly not necessary in general. Again this is clear in
the abelian case $\U(1)^{\times n}$. There, only a whole basis of the algebra
would generate the full algebra (and so $m=n$) yet, $m=1$ is enough as follows
from our discussion in Sec. \ref{sec:3.2.1}: A single displacement $h=-i\vY$
with pairwise incommensurable components (so that $\vk \cdot \vY \neq 0$)
suffices to have an effective action on all Fourier modes simultaneously.

For the general non abelian case the analysis is more complicated so we stick
to our criterion of the set $\{t_k$, \, $k=1,\ldots,m\}$, generating the whole
Lie algebra. Here we find the remarkable result that for semisimple Lie
algebras, $m=2$ seems to be always sufficient.

For instance, for $\SU(2)$ one can take $m=2$, with $t_1 = i\sigma_z$ and
$t_2=i\sigma_x$.\mfootnote{This is not in contradiction with our previous
  remarks around \Eq{5.3}. If $W_{z,n}$ denote the singlet spaces for
  rotations generated by $J_z$ and $J_n$ respectively, \Eq{5.3} expresses the
  condition that $W_z\subset W_n^\perp$. This is more restrictive than $V^j =
  W_z^\perp + W_n^\perp$, $V^j$ being the $2j+1$-dimensional space carrying
  the $\SU(2)$ representation $j$. The condition $W_z\subset W_n^\perp$ does
  require to change $\hat{n}$ for different $j$, whereas $V^j = W_z^\perp +
  W_n^\perp$ does not.} $V^j$ has dimension $2j+1$, the singlet spaces $W^j_1$
and $W^j_2$ have both dimension $1$ for integer $j$ or $0$ for half-integer
$j$.  In both cases $W^{j,\perp}_1 + W^{j,\perp}_2$ fills the space $V^j$. A
canonical prescription to decompose $v = v_1 + v_2$, ~$v\in V^j$,~ $v_k \in
W^{j,\perp}_k$, is to require $v_1=v_2$ along $W^{j,\perp}_1 \cap
W^{j,\perp}_2$. This fixes $v_{1,2}$ uniquely. So a total of $4$ branches
suffice for any complex probability defined on $\SU(2)$.

For $\SU(3)$ the whole algebra is generated by $i\lambda_2$ and $i\lambda_1 +
i\lambda_4$: by taking commutators recursively, eventually a basis of $\su(3)$
is produced. So four branches suffice also in this case.

Moreover, the following two Lie algebra elements seem to generate
the full algebra $\su(N)$ for any $N$:
\begin{equation}\begin{split}
t_1 &= i \, \diag(1,2,\ldots,N-1,-N(N-1)/2)
\\
(t_2)_{\alpha\beta} &= \left\{ \begin{matrix}
i, ~~  \alpha=1, ~\beta\not=1 ~ \text{or} ~
\beta=1, ~\alpha\not=1 
\\
0 ~~ \text{otherwise}
\end{matrix}\right.
.
\end{split}\end{equation}
While we have no rigorous proof of this for all $N$, the statement holds, at
least, for $N \le 8$. In fact, almost any pair of random elements seem to
generate $\su(N)$, and a smaller subalgebra would only be generated by a
careful choice of the pair $(t_1,t_2)$.

The fact that a generic pair of elements $t_{1,2}$ generate the whole algebra
is consistent with $\su(N)$ being simple. As for the direct sum of simple
algebras (semisimple algebras), $m=2$ would hold too. For instance, for $G =
G_1 \times G_1$ with $G_1=\SU(2)$. The algebra has basis
$i\overset{r}{\sigma}_j$, with $j=1,2,3$, and $r=1,2$. It is straightforward
to check that the pair of elements $t_x= i\overset{1}{\sigma}_x + \alpha_x
i\overset{2}{\sigma}_x$ and $t_y= i\overset{1}{\sigma}_y + \alpha_y
i\overset{2}{\sigma}_y$ generates $\su(2)\oplus\su(2)$ for almost any choice
of the real coefficients $\alpha_{x,y}$.

If abelian sectors are added to the semisimple algebra, still $m=2$ is
sufficient to generate the full algebra if the abelian sector is at most
two-dimensional, but not in general. This does not imply though that $m>2$ is
mandatory to fulfill \Eq{4.69}, as already shown for the purely abelian case.

Another remark is that for a higher-dimensional system, with $G =
G_1\times\cdots\times G_1$~($n$ factors), four branches (from $m=2$) may not
be optimal, in the same way that using strictly two-branches (by taking an
irrational $\vY$) is not optimal in the abelian case. Also in the nonabelian
case an uniformity criterion with respect to the $n$ variables is
desirable. The same ideas given in Sec. \ref{sec:3.2.2} apply here, i.e., a
bifurcation for each variable and for each of the $m$ terms. So, the number of
branches changes from $2m$ to $m2^n$. This is illustrated in
Sec. \ref{sec:5.c}.

\subsection{\textsf{Case study I}}

The $\SU(2)$ example discussed in Sec. \ref{sec:4.3} does not contain integer
representations, besides the trivial one, and so the problem of a singular
kernel does not arise.  In order to illustrate the treatment of singular
kernels discussed in the previous subsection, let us consider the following
``complex'' density defined in $\SU(2)$,
\begin{equation}
P(g) = \tr(g)^2
.
\end{equation}
This probability contains components $j=0,1$. It should be noted that actually
$P$ is already real and positive, and normalized, but it needs at least four
branches in the complexified group if one insists on prescribing a certain
decomposition and requires positivity of each component separately.

The density can be written as $P(g) = g^i{}_i \, g^j{}_j$. In order to separate
the trivial representation, we can exploit the relation
\begin{equation}
1 = \det(g) = g^1{}_1 \, g^2{}_2 - g^1{}_2 \, g^2{}_1
,
\end{equation}
to write
\begin{equation}
P(g) = 1  + g^1{}_1 \, g^1{}_1 +  g^2{}_2 \, g^2{}_2 + 
(g^1{}_1 \, g^2{}_2 + g^1{}_2 \, g^2{}_1)
,
\end{equation}
corresponding to the decomposition $P= \sum_{jm} P_{jm}$,
\begin{equation}
P(g) = P_{0,0}  + P_{1,1} + P_{1,-1} + P_{1,0}
.
\end{equation}
The normalization $P_{0,0}=1$ is to be distributed among the three non trivial
components after they are moved into the complexified group manifold.

In a first step we can take a diagonal element $h_z\in \SU(2)_I$,
corresponding to an imaginary rotation
\begin{equation}
h_z = \begin{pmatrix} \omega_z & 0 \\ 0 & \omega_z^{-1} \end{pmatrix}
,\qquad
\omega_z > 1
,
\end{equation}
which would produce [using \Eq{4.54}]
\begin{equation}\begin{split}
\hat{Q}_{1,1}+\hat{Q}_{1,-1} &= 
g^1{}_1 \, g^1{}_1 \, \chi(\omega_z^2)+  g^2{}_2 \, g^2{}_2
\, \chi(\omega_z^{-2}) 
,
\\
\hat{Q}_{1,0}
&=
(g^1{}_1 \, g^2{}_2 + g^1{}_2 \, g^2{}_1) \, \chi(1)
.
\end{split}\end{equation}
The terms $|1,\pm 1\rangle$ can be treated with $h_z$ but $|1,0\rangle$
requires a different transformations since it is invariant under rotations
around the $z$ axis and $\chi(1)$ diverges.

For $P_{1,0}$ one can apply a rotation around the $x$ axis relying on
$\langle 1,0 | \hat{R}(\hat{e}_x,\pi/2) |1,0\rangle = 0$, 
\begin{equation}
h_n = U h_x \, U^{-1}
,
\qquad
h_x = \diag( \omega_x , \omega_x^{-1} )
,
\qquad
U = e^{-i\pi\sigma_y/4}
\,.
\label{eq:5.16}
\end{equation}
An alternative to computing the rank four tensor $\chi(h_n)^{i_1 i_2}{}_{j_1
  j_2}$ is to rotate the elements, as explained in Sec. \ref{sec:4.5}: the
effect of $h_n$ on $g$ corresponds to the action of $h_x$ on $g^\prime =
U^{-1} g \, U$.  Since $h_x$ is diagonal \Eq{4.54} applies. The explicit
result in terms of $g^\prime$ becomes
\begin{equation}\begin{split}
\hat{Q}_{1,0} &=
\frac{1}{2} \big(
(g^\prime{}^1{}_1)^2 \, \chi(\omega_x^2) 
- (g^\prime{}^1{}_2)^2  \, \chi(\omega_x^{-2})
\\ & ~~
- (g^\prime{}^2{}_1)^2  \, \chi(\omega_x^2)
+ (g^\prime{}^2{}_2)^2  \, \chi(\omega_x^{-2})
\big)
.
\end{split}\end{equation}
As advertised no divergence of the type $\chi(1)$ arises.

After this decomposition the expectation values can be expressed through real
weights on the complexified group with four sheets
\begin{equation}
\esp{A}
=
\int_{\SU(2)} dg \, \sum_{\sigma= \pm1} \left(
Q_{z,\sigma}(g) A(g h_z^\sigma) + Q_{x,\sigma}(g) A(g h_n^\sigma)
\right)
.
\end{equation}
Following \Eq{4.53}, here $Q_{z,+}$ is twice the real part of
$\hat{Q}_{1,1}+\hat{Q}_{1,-1}$ plus some constant term $N_{z,+}$ from $P_{0,0}$,
$Q_{z,-}$ is likewise with $h_z^{-1}$, and $Q_{x,\pm}$ likewise for
$\hat{Q}_{1,0}$ with $h_n$. The positive constant terms $N_{z,\pm}$,
$N_{x,\pm}$ add up to one.

In our case, the two functions $Q_{z,\pm}$ turn out to be equal, after choosing
equal normalizations $N_{z,+}= N_{z,-}$, and similarly for $Q_{x,\pm}$. An
explicit calculation gives
\begin{equation}\begin{split}
Q_z(g) &= 
N_z + 2\frac{\cos^2(\psi/2) - \cos^2(\theta)
  \sin^2(\psi/2)}{\omega_z^2 + \omega_z^{-2}}
,
\\
Q_x(g) &= 
N_x + \frac{\cos^2(\theta) + \cos(\psi) \sin^2(\theta) }
{\omega_x^2 + \omega_x^{-2}}
,
\label{eq:5.19}
\end{split}\end{equation}
with
\begin{equation}
N_z,N_x \ge 0 ,
\qquad
2 N_z + 2 N_x  = 1
.
\end{equation}
In the formulas $g = e^{-i\psi \hat{\vpsi}\cdot\sigma/2}$ and $\hat{\vpsi}=
(\theta,\phi)$ in spherical coordinates. $\phi$ does not appear in our case,
related with the invariance of $P(g)$ with respect to similarity
transformations of $g$.

Upon minimization with respect to $(\theta,\psi)$, the conditions ensuring
positive functions $Q_z(g)$ and $Q_x(g)$ are
\begin{equation}\begin{split}
0 \le \min Q_z &= N_z - 2 (\omega_z^2 + \omega_z^{-2})^{-1}
,
\\ 
0 \le \min Q_x &= N_x -  (\omega_x^2 + \omega_x^{-2})^{-1}
.
\end{split}\end{equation}
These inequalities can be fulfilled by taking $\omega_{z,x}$ sufficiently
large. The optimal case (smaller $\omega_{z,x}$) corresponds to $\min Q_z =
\min Q_x = 0$, i.e.,
\begin{equation}\begin{split}
&
2 (\omega_z^2 + \omega_z^{-2})^{-1} = N_z,
\qquad
(\omega_x^2 + \omega_x^{-2})^{-1}  =  \frac{1}{2} - N_z
,
\\ &
0 \le N_z \le \frac{1}{2}
.
\end{split}\end{equation}
For instance, for $N_z=1/3$ one obtains $\omega_z=\omega_x = 1+\sqrt{2}$,
while for $N_z=1/4$, $\omega_z=2.81$ and $\omega_x=1.93$.

Formally it would seem that one could remove, say the two sheets
$Q_{z,\pm}(g)$ by taking $\omega_z\to \infty$ and $\omega_x\to 1$ [or
  $Q_{x,\pm}(g)$ with $\omega_z\to 1.93$ and $\omega_x\to \infty$] however,
this is incorrect. For large $\omega_z$, $Q_{z,\pm}(g)$ is reduced but the
information must be carried by the observable, $A(g h_z^{\pm 1})$. The
observables tend to grow rapidly far from the real manifold producing an
infinite variance in the limit.

It is noteworthy that the functions $Q_{z,x}(g)$ in \Eq{5.19} do not diverge
as $\omega_{z,x} \to 1$. This is a consequence of the fact that our $P(g)$ is
real. In that limit the four distributions have their support on the real
manifold and their sum reproduces the original density:
\begin{equation}
h_z = h_x = 1\,:\quad
2 Q_z(g) + 2 Q_x(g) =  2 (1 + \cos\psi) = P(g)
.
\end{equation}
Even if in the limit $\omega_{z,x}=1$ the sum of the four contributions yield
the original positive density, $Q_z$ and $Q_x$ would not be separately
positive. It is the requirement $Q_z(g) \ge 0$ and $Q_x(g) \ge 0$ that
introduces the non trivial lower bounds on $\omega_z$ and $\omega_x$.

\subsection{\textsf{Case study II}}
\label{sec:5.c}

Next we consider a complex probability defined on $G=\SU(N) \times \SU(N)$,
representing a simplified lattice with two degrees of freedom, namely,
\begin{equation}\begin{split}
P(g_1,g_2) &= \cN^{-1} \left( 1 + \beta \, \tr(g_1^{-1} g_2) \right)
 \left( 1 + \beta \, \tr(g_2^{-1} g_1) \right)
\\& \quad \times 
\tr(g_1) \tr (g_2^{-1})
.
\end{split}\end{equation}
The terms with $\beta$ mimic a gauge action. Those factors are invariant under
$g_i \to \omega^{-1} g_i \omega^\prime$, $i=1,2$,
$\omega,\omega^\prime\in\SU(N)$. The factors $\tr(g_1) \tr(g_2^{-1})$ mimic
Polyakov loops, partially breaking the invariance from $\SU(N) \times \SU(N)$
to $\SU(N)$ ($\omega=\omega^\prime$), but preserving global center invariance,
$g_i\to z g_i$, $z\in\U(1)$, $z^n=1$.

For $N>2$ the normalization of $P(g)$ comes solely from $\tr(g_1^{-1}
g_2)\tr(g_1) \tr (g_2^{-1})$, however when $N=2$ the term $\tr(g_2^{-1}
g_1)\tr(g_1) \tr (g_2^{-1})$ gives an identical contribution, due to
$\tr(g^{-1})=\tr(g)$. Thus $P(g)$ is normalized with\mfootnote{Using standard
  $\SU(N)$ group integration rules \cite{Creutz:1984mg}.}
\begin{equation}
\cN =  \left\{ \begin{matrix}
 \beta
&  \quad (N=2) \\
 \beta/N
&  \quad (N>2) \\
\end{matrix} \right.
.
\end{equation}

One can decompose $P(g)$ in monomials, as in \Eq{4.60}, and apply a diagonal
element of $G_I$, $h_z$, with parameters $\overset{r}{\omega}_{z,i} > 0$,
$r=1,2$, $i=1,\ldots,N$. The complex representation $\hat{Q}$ is then obtained
as in \Eq{4.61}. Each term in $\hat{Q}$ picks up a factor $\chi(\Omega_z)$ and
the problem of singular kernels corresponds to the components for which
$\Omega_z=1$. Such components should be treated with a different element $h_n$
of $G_I$.

We can see that the terms which are singular under $h_z$, i.e., contain the
trivial representation (in a reduction with respect to the subgroup generated
by $h_z$) are contained in $\tr(g_1^{-1} g_2)\tr(g_1) \tr (g_2^{-1})$.
\begin{equation}\begin{split}
\tr(g_1^{-1} g_2) & \tr(g_1) \tr (g_2^{-1})
=
\overset{1}{g}{}^{-1}{}^i{}_j \, \overset{2}{g}{}^j{}_i \,
\overset{1}{g}{}^k{}_k \, \overset{2}{g}{}^{-1}{}^\ell{}_\ell
,\qquad
\\
\Omega_z &= \overset{1}{\omega}{}^{-1}_{z,i} \,\overset{2}{\omega}{}_{z,i}
\,\overset{1}{\omega}{}_{z,k} \,\overset{2}{\omega}{}^{-1}_{z,\ell} 
\,.
\end{split}\end{equation}
Generically $\Omega_z=1$ when $i=k=\ell$, a total of $N^2$ terms:
\begin{equation}
\overset{1}{g}{}^i{}_i \, 
\overset{1}{g}{}^{-1}{}^i{}_j \, 
\overset{2}{g}{}^j{}_i \,
\overset{2}{g}{}^{-1}{}^i{}_i
,\qquad
i,j=1,\ldots,N
,
\qquad \Omega_z=1
\,.
\label{eq:5.26}
\end{equation}
In order to choose $h_n$, this can be analyzed as follows.  Each factor
$\overset{r}{g} \otimes \overset{r}{g}{}^{-1}$, $r=1,2$, can be reduced as
trivial plus adjoint representation and contains $N$ singlets under a diagonal
$\overset{r}{h}$ (one from the trivial representation and $N-1$ from the
adjoint). This $N$-dimensional space is spanned by the $N\times N$ diagonal
matrices (the traceless matrices being in the adjoint sector). Therefore, out
of the $N^2$ singular terms, one comes from the trivial representation of
$\SU(N) \times \SU(N)$ and the remaining $N^2-1$ come from the adjoint
representation in one or both factors. So $h_n$ can be chosen in the form
$\overset{1}{h}{}_n\overset{2}{h}{}_n$ with the condition that
$\overset{r}{h}{}_n$ must act {\em effectively} on the components of
$\overset{r}{g} \otimes \overset{r}{g}{}^{-1}$ which are invariant under
$h_z$. If $h_n$ is written as $U h_x U^{-1}$, with diagonal $h_x$, $U$ must be
chosen so that any traceless diagonal matrix, upon rotation by $U$, has not
overlap with any other traceless diagonal matrix (similar to the condition in
\Eq{5.3}):
\begin{equation}\begin{split}
0 &= \tr( A_z U A_x U^{-1})
,
\quad U\in \SU(N)
\\ &
\quad \text{for all~ $A_{z,x}$ traceless and diagonal}
.
\end{split}\end{equation}
An easy calculation shows that this implies
\begin{equation}
 |U^j{}_\ell|^2 = \frac{1}{N}
\qquad j,\ell = 1,\ldots, N
,
\label{eq:5.28}
\end{equation}
and an explicit solution is
\begin{equation}
 U^j{}_\ell = \frac{1}{\sqrt{N}} e^{i2\pi (j-1)(\ell-1)/N} 
\qquad j,\ell = 1,\ldots, N
.
\end{equation}
In particular for $N=2$, $U= e^{-i\pi\sigma_y/4}$ [consistently with
  \Eq{5.16}].

One can now verify that the previously singular terms of \Eq{5.26} are not
singular under $h_n$ [upon removing the trivial representation of
$\SU(N)\times \SU(N)$]. To do that we use
\begin{equation}
g^i{}_j
\mapsto (g h_n)^i{}_j
= (g U h_x U^{-1}){}^i{}_j
= \sum_\ell (g U)^i{}_\ell  \omega_{x,\ell} (U^{-1})^\ell{}_j
.
\end{equation}
It is sufficient to consider just one of the factors in \nec{5.26}:
\begin{equation}\begin{split}
\overset{1}{g}{}^i{}_i \, 
\overset{1}{g}{}^{-1}{}^i{}_j \, 
&\mapsto
\sum_\ell
(\overset{1}{g} U)^i{}_\ell  \overset{1}{\omega}_{x,\ell} (U^{-1}){}^\ell{}_i
\sum_m
U{}^i{}_m
\overset{1}{\omega}{}_{x,m}^{-1}
(U^{-1}\overset{1}{g}{}^{-1})^m{}_j 
,
\\ & 
\qquad
\overset{1}{\Omega}_x = \overset{1}{\omega}_{x,\ell}
\overset{1}{\omega}{}_{x,m}^{-1}
.
\end{split}\end{equation}
The possible singular contributions, $\overset{1}{\Omega}_x=1$, would come
from $\ell=m$. For these terms one obtains
\begin{equation}
\sum_\ell
(\overset{1}{g} U)^i{}_\ell  (U^{-1}){}^\ell{}_i
U{}^i{}_\ell
(U^{-1}\overset{1}{g}{}^{-1})^\ell{}_j 
= \frac{1}{N}\delta^i{}_j
\,,
\label{eq:5.33}
\end{equation}
using \Eq{5.28}.\mfootnote{Alternatively, one can derive the condition in
  \Eq{5.28} by requiring the fulfillment of \nec{5.33}.} An identical result
is obtained for the second factor $\overset{2}{g}{}^j{}_i \,
\overset{2}{g}{}^{-1}{}^i{}_i$. So the terms that remain invariant under $h_n$
are
\begin{equation}
\overset{1}{g}{}^i{}_i \, 
\overset{1}{g}{}^{-1}{}^i{}_j \, 
\overset{2}{g}{}^j{}_i \,
\overset{2}{g}{}^{-1}{}^i{}_i
\quad
\mapsto
\quad 
\frac{1}{N}\delta^i{}_j
\frac{1}{N}\delta^j{}_i = \frac{1}{N}
.
\end{equation}
This is independent of $g$ and corresponds to the trivial representation of the
full group, which always has to be extracted from $P(g)$. The trivial
representation saturates the normalization, and indeed, the final result $1/N$
combined with the factor $\cN^{-1}\beta$ (or $2\cN^{-1}\beta$ for $N=2$)
checks that $P(g)$ is normalized.

After extraction of the constant mode, $P(g)$ can be written as a sum of two
terms, namely, the monomials to be rotated with $h_z$ and those to be rotated
with $h_n$,
\begin{equation}
P(g) = 1 + P_z(g) + P_x(g)
\,.
\end{equation}
It should be noted that $P_z$ (the same goes for $P_x$) is non singular for
generic values of $\overset{r}{\omega}_{z,j}$, but new divergences can appear
for especial correlated values. For instance a term with
$\chi(\overset{1}{\omega}{}_{z,i} \overset{2}{\omega}{}_{z,j}^{-1})$ prevents
taking these two $\omega$'s to be equal.

Let us consider the case $N=2$ in more detail:
\begin{equation}\begin{split}
P(g) &= \frac{1}{\beta}
\left( 1 + \beta \, \tr(g_1^{-1} g_2) \right)^2
 \tr(g_1) \tr (g_2)
,
\\
g &= (g_1,g_2) \in \SU(2) \times \SU(2)
.
\end{split}\end{equation}
In addition, for simplicity, we will assume $\beta>0$. 

The complex representations associated to the two sectors $P_z$ and $P_x$ are
easily obtained using $\overset{r}{h}_z =
\diag(\overset{r}{\omega}_z,\overset{r}{\omega}{}_z^{-1})$, and similarly for
$h_x$. This gives [expanding $P_{z,x}$ in monomials and applying \Eq{4.54}]
\begin{equation}\begin{split}
\hat{Q}_z &= 2\,
\overset{1}{g}{}^1{}_1 \, \overset{2}{g}{}^1{}_1 
\left(
\overset{1}{g}{}^1{}_1 \, \overset{2}{g}{}^2{}_2
- \overset{1}{g}{}^2{}_1 \, \overset{2}{g}{}^1{}_2
\right)
\chi(\overset{1}{\omega}{}_z^2)
+ \cdots
\quad\text{(16 terms)}
\\
\hat{Q}_x &=
\frac{1}{2}\,
\left( \overset{1}{g}{}^1{}_1 + \overset{1}{g}{}^1{}_2 \right)
\left( \overset{1}{g}{}^2{}_1 + \overset{1}{g}{}^2{}_2 \right)
\chi(\overset{1}{\omega}{}_x^2)
+ \cdots
\quad~\,\text{(8 terms)}
\end{split}\end{equation}
The 16 terms in $\hat{Q}_z$ are classified by 16 combinations of the exponents
$(k,m)$ in $\chi(\overset{1}{\omega}{}_z^k \overset{2}{\omega}{}_z^m )$, and
similarly for the 8 terms in the $x$ sector.

Taking real parts, and changing $\omega \to \omega^{-1}$, for the various
$\omega$, produces the four distributions corresponding to four sheets on the
complexified group, two sheets for each sector $z$ and $x$.  After this step
the dependence on the $\Omega_z$'s is through the symmetric combination
$\chi(\Omega_z ) + \chi(\Omega_z^{-1} )$, and similarly in the $x$
sector. This feature is an idiosyncrasy of this complex probability and group.

However, as discussed in Sec. \ref{sec:3.2.2}, instead of two sheets, it is
preferable to use $2^n$ sheets for $n$ variables, $n=2$ in our case. This
allows to take the same $\overset{r}\omega_z$ for $r=1$ and $r=2$ (and
similarly for $\overset{r}\omega_x$), and also to reduce the numerical value
of the $\Omega_{z,x}$ required to have positive distributions.

The method is explained in Sec. \ref{sec:3.2.2}: Initially there are two
sheets in the $z$ sector (everything is similar in the $x$ sector), produced
by the transformations $(\overset{1}{\omega}{}_z,\overset{2}{\omega}{}_z)$ and
$(\overset{1}{\omega}{}_z^{-1},\overset{2}{\omega}{}_z^{-1})$. Then a term
with $\Omega_z = \overset{1}{\omega}{}_z^k \overset{2}{\omega}{}_z^m $ is
unchanged if $km>0$. If $km<0$, it is changed to $\overset{1}{\omega}{}_z^k
\overset{2}{\omega}{}_z^{-m}$ and moved to the sheet
$(\overset{1}{\omega}{}_z,\overset{2}{\omega}{}_z^{-1})$. When $km=0$, half of
the term stays and the other half is moved to the opposite
sheet.\mfootnote{The coordinate that is reflected is that with a zeroth power
  in $\Omega_z$.}

Following this procedure eight branches, with functions $Q_z{}_{\pm,\pm}(g)$
and $Q_x{}_{\pm,\pm}(g)$, are obtained. Taking the symmetric choice
\begin{equation}
 \overset{1}{\omega}{}_z = \overset{2}{\omega}{}_z \equiv \omega_z,
\qquad
 \overset{1}{\omega}{}_x = \overset{2}{\omega}{}_x \equiv \omega_x
,
\end{equation}
$Q_z{}_{++}$ contains terms $\Omega_z= \omega_z^m$ with $m=2 $, while
$Q_z{}_{+-}$ has $m=2,4,6$. In the $x$ sector, $Q_x{}_{++}$ and
$Q_x{}_{+-}$ both contain terms $\Omega_x = \omega_x^m$ with $m= 2,4$.

In order to apply the method, the unit normalization of $P$ must be
distributed among the eight branches to produce positive distributions. To
achieve this $\omega_{z,x}$ have to be taken sufficiently large so that all
minima of $Q_{z,\pm,\pm} $ and $Q_{x,\pm,\pm}$, and their sums, are above
$-1$.\mfootnote{Here the functions $Q_{z,x}$ do not contain the constant
  modes. The conditions to be above $-1$ are similar to those in
  \Eq{3.10a}. They guarantee that a global unit normalization can be added to
  the various branches in the form constant modes to make these functions
  positive.} The minima of these functions (over the manifold
$\SU(2)\times\SU(2)$) will depend on the choice of $\omega_{z,x}$ and $\beta$
and presumably they cannot be found in a closed analytic form. Our approach
has been to split the functions into a sum of terms classified by their
dependence on $\Omega$ and its power of $\beta$ and (numerically) find an
independent minimum for each such term. This provides a lower bound to the
true minimum, since there can be cancellations between terms which are
neglected in our approach. A lower bound is sufficient for our purposes. The
lower bounds to the minima so obtained are
\begin{equation}\begin{split}
\min Q_{z,++} & \ge -(4\beta + 4 + \frac{2}{\beta})\chi_s(\omega_z^2)
\\
\min Q_{z,+-} & \ge 
-( 6 \beta + 4 + \frac{2}{\beta} )\chi_s(\omega_z^2)
\\&\quad
-( 4 \beta +  4 )\chi_s(\omega_z^4)
-2\beta \chi_s(\omega_z^6)
\\
\min Q_{x,++} & \ge -\chi_s(\omega_x^2) - \chi_s(\omega_x^4)\\
\min Q_{x,+-} & \ge -\chi_s(\omega_x^2) - \chi_s(\omega_x^4)
\end{split}\end{equation}
where $\beta>0$ and
\begin{equation}
\chi_s(\Omega) \equiv
\chi(\Omega)+\chi(\Omega^{-1}) = \frac{1}{\Omega+\Omega^{-1}}
.
\label{eq:5.39}
\end{equation}
It is noteworthy that the coefficients found numerically turn out to be simple
numbers.  Remarkably, choosing concrete values of $\beta$ (to combine various
terms and so increase the minimum) has not resulted in any improvement. So the
method used seems to be numerically accurate, producing good estimates for the
minima.

Since all expressions in \Eq{5.39} are negative, it is sufficient to
constrain their sum. The optimal values of the pair $(\omega_z,\omega_x)$ are
thus constrained by the condition
\begin{equation}\begin{split}
\frac{1}{2} &= (10\beta + 8 + \frac{4}{\beta})\chi_s(\omega_z^2) +( 4 \beta + 4
)\chi_s(\omega_z^4) 
\\&\quad
+2\beta \chi_s(\omega_z^6) + 2\chi_s(\omega_x^2) +2
\chi_s(\omega_x^4).
\end{split}\end{equation}
Saturation of the equality by the terms with $\omega_z$ (by letting $\omega_x$
to be as high as needed) yields the bounds $\omega_z\ge e^{1.90}$ for $\beta=1$
and $\omega_z\ge e^{2.05}$ for $\beta=2$. Likewise $\omega_x\ge e^{0.72}$ for
any value of $\beta$.

The choice $\omega_z = \omega_x $, for $\beta=1$ and $\beta=2$ gives
$\omega_z= 6.95= e^{1.94}$ and $\omega_z=8.02=e^{2.08}$, respectively.  In
this scenario most of the normalization ($92\%$) goes to the $z$-sheets, with
$N_{z,++}=0.207$, $N_{z,+-}=0.251$, and $N_{x,++}=N_{x,+-}=0.021$.  Using
these parameters, we have analyzed a sample operator, $\cO = \tr(g^{-1}_1
g_2)$, with exact expectation value is $\esp{\cO} = \beta + 1/(2\beta)$. In
our representation, the expectation value comes only from the sheets
$Q_{z,+-}$ and $Q_{z,-+}$, the other sheets giving a vanishing
contribution. All the sheets contribute to the variance, which can be computed
analytically, but $Q_{z,+-}$ and $Q_{z,-+}$ are also dominant for the
variance, through a large $\beta$-independent term, namely, $N_{z,+-}
\omega_z^4$. For $\beta=1$ the total variance is $634.$ This number
  depends also on the precise definition of the variance. The number quoted
  refers to the variance knowing the normalization of each branch. If this
  were not known one should add the variance of the means on each branch
  around the total mean. This extra variance is a comparatively small number
  in our case, $2.22$ for $\beta=1$.

This is to be compared with the variance obtained using simple reweighting
with $|P(g)|$ (and assuming that its normalization is known). This variance
can be obtained analytically, obtaining
\begin{equation}
{\rm Var}_{{\rm RW}} = 0.878 + 0.374 \beta^2 + \frac{0.331}{\beta^2}
.
\end{equation}
This gives number of the order of unity for $\beta$'s of the same
order. Therefore in this case reweighting has a much better performance than
the representation, however such good performance should deteriorate
exponentially with the number of variables.


\section{\textsf{Summary and conclusions}
\label{sec:6}}

In this work we have analyzed the problem of constructing representations of
complex weights within the two-branch approach, which is probably optimal from
the point of view of localization. In this regard, new localization conditions
on positive representations have been uncovered in Sec. \ref{sec:2.c}
(cf. \Eq{2.4a}).

In the abelian many-dimensional case a solution is found
(Sec. \ref{sec:3.2.2}) to the problem of treating all variables on an equal
footing, and simplifying the choice of parameters. The method proposed is to
share the weight over $2^n$ sheets, for $n$ variables. This allows to use
copies of the real manifold which are closer (to the real manifold), and so
with smaller variance.

The other main novelty is the study of representations of complex weight
defined on compact group manifolds, within a two-branch approach
(Sec. \ref{sec:4}). In this scheme two copies of the (real) group are obtained
upon translation by an imaginary element and its inverse. Each copy carries a
positive distribution whose analytic continuation, when added, reproduces the
original complex weight. The construction is illustrated in detail for a
complex weight defined on $\SU(2)$. When the imaginary element does not act
effectively on some of the components of the complex weight, so that they are
not moved to the complexified group manifold, an obstruction is met in the
form of a singular kernel. We have shown (Sec. \ref{sec:5.a}) how the
obstruction can be removed, namely, by decomposing the complex weight into
components, each of which can be acted effectively by some imaginary
element. We have shown that such a decomposition always exists.

Explicit examples have been worked out for $\SU(2)$ with integer spin
representations (hence, subject to obstruction) and for $\SU(2) \times
\SU(2)$, also presenting singular kernels.

While the abelian case had been considered earlier, no explicit construction
of positive representations existed for nonabelian groups in the literature,
and indeed unexpected impediments have had to be sorted out. In view of
this, in general (an exception being Sec. \ref{sec:III.D.1}) in this
exploratory work we have not aimed at a rigorous mathematical formulation
(specifying precise domains of definitions, norms, etc). However there are no
foreseeable obstructions to such a treatment for complex densities $P(g)$
which are distributions defined on compact groups and involving just a finite
number of irreducible representations of the group. Much more challenging
should be the rigorous mathematical treatment for more general complex
densities, depending on how much generality is allowed.

An interesting lesson from the direct representation approach to the sign
problem is that even realistic theories like lattice QCD with a chemical
potential must admit such representations, however complicated and nonlocal
they might be. This opens the possibility of trying to directly model a local
and positive action on the complexified manifold, incorporating the chemical
potential, and hopefully in the same universality class as the original QCD
problem.

This study was motivated by the sign problem. A natural question is the
practical application of this study to addressing this difficult problem. The
type of direct representation approach considered here (as opposed to say
Complex Langevin, where $\rho$ is never explicitly constructed) can shed light
on aspects and general properties of the representation problem, including the
crucial issue of localization. As noted in Sec \ref{sec:2.c} such analysis can
show for instance that for certain $P(\vx)$ Complex Langevin will not converge
to the right distribution, even without carrying out a detailed stochastic
simulation. However, it should be clear that a naive direct approach cannot
provide a straightforward solution to the sign problem. The reason is simple
enough: to reconstruct the positive representation $\rho$ one needs the
Fourier modes $\tilde{P}_\vk$ (taking an abelian periodic setting, for
definiteness) but these are just the expectation values $\esp{e^{-i\vk\vx}}_P$
and obtaining them was precisely the whole point of the Monte Carlo
calculation.

Nevertheless, this does not mean that the interest of the direct
representations, as those considered here, must remain at a merely theoretical
level only. It is a common place that when new ideas, even purely theoretical
ones, are examined and the results are assimilated, there is always a chance
to eventually make practical use of them (often in combination of other
existing ideas) employing some ingenuity which a priori cannot be foreseen. In
our case there are in fact routes to practical applications of the ideas
presented here. The most promising one is through a complex version of the
Gibbs sampling. This possibility has been investigated in
\cite{Salcedo:2015jxd} and further analyzed in \cite{Salcedo:2017tnt}.  In the
standard Gibbs sampling or heat bath method each variable (or site in a
lattice problem) is updated in turn using as distribution the conditional
probability of the variable with respect to the other ones, which act as a
background. In practice, the actions being local, only a small number of
neighboring sites are involved in the update. In the complex version, the
procedure is analogous, a site is updated in the complexified manifold using a
{\em positive representation} of the conditional probability with respect to
the neighboring variables, which lie themselves on the complexified manifold.
The interest of this approach should be clear: the main problem of a global
direct representation is the construction of the positive representation
itself for the whole system, however, in the heat bath method only a single
variable is treated at each step, and in this case it is relatively simple to
construct the required positive representation. The required expectation
values can be computed, for instance, through direct numerical quadrature
methods, or other. Certainly, each update will be costly, but it is also true
that the sign problem is a hard one. Besides, it should be possible, with some
skillfulness to construct parameterizations of the positive representations to
alleviate the representation construction problem. The method has been applied
in \cite{Salcedo:2015jxd} in detail to simple complex actions of scalar fields
for relatively large lattices. It was found that the approach works, providing
non trivial results, for moderate values of the complex coupling constant, but
becomes unstable for large values. As discussed in \cite{Salcedo:2017tnt} an
important limitation of the complex Gibbs method is the possible presence of
zeroes on the complexified manifold in the marginal probabilities, since they
appear as a denominator in the conditional probability to be
represented. Those zeroes introduce singularities in the form of poles, in
such a way that effectively one is dealing with observables which are not
holomorphic, spoiling the validity of the representation relation \Eq{2.2}.
On the other hand, the presence or not of such zeroes can be monitored during
the Monte Carlo simulation, which allows to asses the accuracy of the
calculation. A possible way out to the problem of marginal zeroes could be to
use deformed two-branch manifolds avoiding the regions with such zeroes. As
noted at the end of Sec. \ref{sec:3.1} such deformations are possible but the
construction of positives representation becomes harder, and most importantly,
the localization of the zeroes may make them impossible or very difficult to
avoid. Another possible route to the use of the direct representations is
through the convolution formulas, such as \nec{3.20} or \nec{4.29}. Somehow
one would have to sample the real and positive function $N+\Re(C*P)$, however
the way to do this is much more speculative and may be it would not be simpler
than the original problem. On the other hand, in favor of the idea that such
formulas could be use for sampling is the fact that the method in \nec{2.12}
is nothing but a convolution, which in fact does not require an explicit
construction of $\rho(\vz)$ explicitly. While the method in \nec{2.12} is by
no means optimal it shows that using additional input (the function
$P_0(\vx)$, etc) sampling through convolutions is possible.

\acknowledgments I thank E. Seiler and J. Wosiek for discussions. This work
has been partially supported by the Spanish MINECO (grants Nos.
FIS2014-59386-P and FIS2017-85053-C2-1-P) and by the Junta de Andalucia (grant
No. FQM-225).

\appendix
\section{\textsf{Proof of \Eq{2.8}}
\label{app:A}}

We want to show that the positive distribution $q_\vh(\vz)$ in \nec{2.8} is a
representation of $Q_\vh(\vx)$ in \nec{2.7}. Let $A(\vz)$ be an entire
holomorphic observable which we assume to be exponentially bounded. Then
\begin{equation}\begin{split}
\esp{A(\vz)}_{q_\vh} &=
\int A(\vz) \, q_\vh(\vz)\,d^{2n}z
\\& =
\int A(\vz) \, q(\zeta)\delta(\vz-\zeta \vh) \,d^2\zeta\,d^{2n}z
\\& =
\int A(\zeta \vh)  \, q(\zeta) \,d^2\zeta
=
\int A(x \vh) Q(x) \,dx
\\& =
\int A(x \vh) (\delta(x) + \delta'(x)) \,dx
\\& =
A(\vcero) - \vh\cdot\vnabla A(\vcero)
=
\int A(\vx) Q_\vh(\vx)\,d^nx
\\& =
\esp{A(\vx)}_{Q_\vh}
\,.
\end{split}\end{equation}

\end{document}